\documentclass[%
 reprint,
 floats,
nofootinbib,
 amsmath,amssymb,
 aps,
 prc
]{revtex4-2}

\usepackage[T1]{fontenc}

\usepackage{amsmath,
            hyperref,
            url,
            graphicx,
            color}
           
\let\Re\relax
\let\Im\relax
\DeclareMathOperator\Re{Re}
\DeclareMathOperator\Im{Im}

\newcommand\unit[1]{~\mathrm{#1}}
\newcommand\MeV[1]{#1\unit{MeV}}

\newcommand\op[1]{\hat{#1}}
\newcommand\mat[1]{#1}

\newcommand\of[1]{\left(#1\right)}

\newcommand\wrt[1]{\mathrm{d}#1\,}

\newcommand\bra[1]{\left<#1\right|}
\newcommand\ket[1]{\left|#1\right>}
\newcommand\braket[2]{\left<#1\middle|#2\right>}
\newcommand\braopket[3]{\left<#1\middle|#2\middle|#3\right>}
\newcommand\expect[1]{\left<#1\right>}

\newcommand\deltaf[1]{\delta\of{#1}}

\newcommand\comm[2]{\left[#1, #2\right]}
\newcommand\acomm[2]{\left\{#1, #2\right\}}

\newcommand\hconj[1]{{#1}^\dagger}

\newcommand\conj[1]{{#1}^*}

\newcommand\expp[1]{\mathrm{e}^{#1}}

\DeclareMathOperator\pf{pf}

\newcommand\T[1]{{#1}^\mathrm{T}}

\newcommand\inv[1]{{#1}^{-1}}

\newcommand\id{\mathrm{I}}

\newcommand\isotope[2]{{}^{#1}\mathrm{#2}}
\newcommand\mg[1]{\isotope{#1}{Mg}}

\newcommand\Nmax{\ensuremath{N_{max}}}
\newcommand\Ncut{\ensuremath{N_{cut}}}
\newcommand\ncut{\ensuremath{n_{cut}}}
\newcommand\ngrid{\ensuremath{n_{grid}}}
\newcommand\temp{\ensuremath{b}}
\newcommand\qnum{\ensuremath{q_{num}}}
\newcommand\sig{\ensuremath{r_x}}

\newcommand\mkplot[4][0.9\textwidth]{%
    \begin{figure}[ht]%
        \centering%
        \includegraphics[width=#1]{#2/#3}%
        \caption{#4}%
        \label{fig:#3}%
    \end{figure}}

\usepackage{xfrac}

\begin{document}


\title{Microscopic optical potential framework applied to neutron scattering on deformed $^{48,50}$Cr}

\author{J. Bostr\"om}
\affiliation{Division of Mathematical Physics, Department of Physics, LTH, \\ Lund University, PO Box 118, S-22100 Lund, Sweden}
\author{B. G. Carlsson}
\affiliation{Division of Mathematical Physics, Department of Physics, LTH, \\ Lund University, PO Box 118, S-22100 Lund, Sweden}
\author{A. Idini} 
\affiliation{Division of Mathematical Physics, Department of Physics, LTH, \\ Lund University, PO Box 118, S-22100 Lund, Sweden}

\begin{abstract}
    We formulate and implement a microscopic framework to derive an optical potential from the solution to an effective Hamiltonian and use it to calculate neutron scattering cross sections for the deformed nuclei $^{24}$Mg, $^{48}$Cr and $^{50}$Cr. 
    This approach is based on a symmetry-restored multi-excitation generator coordinate method (GCM), enabling the consistent treatment of both nuclear structure and reaction observables. Through this method, non-local optical potentials corresponding to a Hamiltonian can potentially be constructed for any nucleus in the whole nuclide chart. We use this to perform reaction calculations employing quadrupole deformed triaxial configurations, obtaining results for $A\approx 50$ chromium isotopes, and study the properties of the calculated non-local optical potentials. This work further advances the unified treatment of structure and reaction, within a framework that exploits the intrinsic symmetries of nuclei.
\end{abstract}

\maketitle

\section{Introduction}





Nuclear reactions are one of the most important tools for studying atomic nuclei, and play a large role in several astrophysical processes, as well as in nuclear reactors.
In reactors, Chromium isotopes are of particular interest since they are important components of stainless steel.
Neutron cross sections for $\isotope{50}{Cr}$ are included in the Nuclear Energy Agency (NEA) high priority request list \cite{Dupont:20} in order to improve the current uncertainties of cross section measurements. Neutron reaction cross sections are the topic of multiple recent and planned experiments \cite{Guerrero:13,PerezMaroto:25}.

One of the most important methods for studying nuclear reactions is the optical potential, 
a method which employs an effective potential that describes particle scattering while incorporating the effects of all other reaction channels as well \cite{Johnson:20}.
Optical potentials are routinely used to model nuclear reactions in many contexts, for example the analysis of experiments in rare isotope beams \cite{Hebborn:23,Crawford:24}, modelling of stellar burning and nucleosynthesis \cite{Nunes:05,Schatz:22}, novel nuclear reactor designs \cite{Rochman:25},
and other applications \cite{Hebborn:23}.

The most widely used optical potentials are phenomenological, that is fitted to experimental data \cite{Koning:03,Pruitt:23}.
This limits their predictive power for nuclei far away from the fitting region, such as for radioactive beams and exotic nuclei present in astrophysical processes, and for energies not usually probed by experiments.
However, optical potentials are in principle an exact method that can be constructed from nuclear structure wavefunctions \cite{Feshbach:58a,Feshbach:58b}, nonetheless it is still an open problem which has been the subject of extensive research.
Recent efforts have been made to create optical potentials based on a number of different many--body methods,
like coupled--cluster method \cite{Rotureau:17}, symmetry adapted no-core shell model \cite{Burrows:24}, self--consistent Green's functions \cite{Idini:19}, nuclear structure method \cite{Blanchon:15}, and the first application of the method described in depth in this article \cite{Bostrom:25}.

At the same time, progress has been made on using multi-reference methods for describing nuclear structure.
These methods use multiple reference states to construct the basis in which the system Hamiltonian is solved, allowing collective properties to be captured by the basis.
Advances in multi-reference methods include many-body methods
such as the generator coordinate method (GCM) \cite{Ljungberg:22,Lin:24}, perturbative GCM based on in-medium similarity renormalization group \cite{Duguet:23,Frosini:22}, and tailored coupled cluster \cite{Sun:23}.
Symmetry breaking and restoration by projection is another way to incorporate collective correlations,
and have been recently performed for single-reference coupled cluster \cite{Sun:25}.

A large part of the computational cost arises from the treatment of the Hamiltonian, and especially interactions based on chiral effective field theory are highly limited by the computation of Hamiltonian matrix elements in large model spaces, particularly three body forces.
Efforts have therefore been made to create Hamiltonians of reduced ranks, for example using tensor decompositions and low-momentum interactions \cite{Tichai:19,Bogner:10,Plies:25}. Another strategy is to build effective Hamiltonians based on pairing and multipole interaction terms, which offer a general compact representation of the nuclear interaction and yield matrix elements that can be evaluated efficiently \cite{Idini:12, Idini:15, Broglia:16, Ljungberg:22, Ljungberg:23proc}.

To apply microscopic structure methods to scattering,
the interaction between the bound states and
the scattering continuum also needs to be considered in a consistent way.
To capture the effects of the continuum using a finite basis,
multiple approaches have been developed,
such as the Green's function formalism \cite{Dickhoff:04, Broglia:16, Idini:19}, Gamow shell model \cite{Jaganathen:14,Mercenne:19},
or the R-matrix \cite{Descouvemont:10} and J-matrix \cite{Alhaidari:08,Shirokov:18} methods.
In \cite{Bostrom:25}, the Green's function formalism was used to calculate an optical potential for $\isotope{24}{Mg}$ from the result of a GCM calculation, and this method is further elaborated in this article. Here it is applied to $\isotope{48}{Cr}$ and $\isotope{50}{Cr}$,
while $\isotope{24}{Mg}$ is studied in detail in Appendix \ref{sec:convergence}, \ref{sec:completion} and \ref{sec:imag_part}.



The GCM is a general method to find solution to a Hamiltonian exploiting the principle of symmetry breaking.
It consists of constructing a set of nonorthogonal wavefunctions parameterized by so called generator coordinates, which describe the most important degrees of freedom of the nucleus, such as its deformation, angular momentum, and amount of pairing.
By explicitly breaking symmetries,
static correlations can be captured,
and the correct symmetries of the Hamiltonian
are then restored using projection.
The initial wavefunctions can be constructed by minimizing the energy of Hartree-Fock-Bogoliubov (HFB) states with constraints,
and through recent advancements in the formalism there are now exact expressions for overlaps and Hamiltonian matrix elements valid for the large model spaces necessary for scattering calculations \cite{Carlsson:21,Bostrom:23proc}.
In this article, the results of GCM calculations are used to construct scattering potentials, using the solution states of the $A\pm1$ systems and their corresponding spectroscopic amplitudes.

The Hamiltonian used is an effective Hamiltonian that reproduce the results of a deformed density functional theory (DFT) calculation, consisting of a spherical one-body part, a quadrupole-quadrupole-like term, and a pairing term \cite{Ljungberg:22}. A set of states is first found by performing a number of HFB calculations with constraints. The particle number and angular momentum quantum numbers are then restored through projection, the Hamiltonian and overlap matrices are calculated, and the resulting Hill-Wheeler equation is solved.
The solutions are then used to construct an optical potential by also calculating the spectroscopic amplitudes. Since the solution set is not complete,
sum rules are used to estimate an averaged contribution from the states not found. Because of this, the potential construction can be generalized to any many body solution method capable of calculating energies and spectroscopic amplitudes.

The results of this work indicate that optical potentials can be derived directly from microscopic Hamiltonians, achieving high--quality agreement with data without case-by-case phenomenological tuning. This has important implications for modelling reactions in regimes where standard optical potentials are insufficient.
Moreover, reaction observables have recently been recognized as essential constraints on nuclear structure itself \cite{Idini:20, Bertulani:21, Idini:23proc, Idini:23proc2}. Consequently, a unified and internally consistent description of nuclear structure and reactions has the potential to advance nuclear modelling beyond the mere reproduction of reaction observables.

In the following, an overview of the many body method is first presented in \autoref{sec:framework}, together with the construction of the optical potential, the sum rules, the process of using sum rules to get a complete basis, and finally the calculation of scattering observables.
The framework is then applied to the case of neutron scattering on $\isotope{48}{Cr}$ and $\isotope{50}{Cr}$ in \autoref{sec:Cr}. Average properties of the obtained optical potential are then analysed in \autoref{sec:pot_props}. Finally, concluding remarks and future outlook of this research are discussed in \autoref{sec:conclusion}. Note that convergence is discussed in Appendix \ref{sec:convergence}, together with discussion and analysis on the completion method and imaginary part in Appendix \ref{sec:completion} and \ref{sec:imag_part} respectively.

\section{Framework}\label{sec:framework}

\subsection{Many body method}

To construct an optical potential,
any many body method capable of calculating an energy spectrum
and corresponding spectroscopic amplitudes can be used
with the prescriptions proposed in this framework. 
For the calculations performed in this article,
the GCM was used with an effective Hamiltonian,
previously outlined in \cite{Ljungberg:22,Ljungberg:24}.

The GCM describes the many body states using a set of
wavefunctions $\ket{\phi\of{\vec{q}}}$ which are constructed from
the generator coordinates $\vec{q}$.
Practically, a discrete set of coordinates $\left\{\vec{q}_i\right\}$
is chosen, which gives a set of basis states
$\left\{\ket{\phi\of{\vec{q}_i}}\right\}$.
To describe collective behavior, these states are selected so that symmetries
of the original Hamiltonian are explicitly broken,
and are then restored by projection.

Angular momentum is restored by the corresponding projection operator
\begin{align}
    \op{P}^I_{M,K} = \frac{2I+1}{16\pi^2}\int_0^{2\pi}\int_0^\pi\int_0^{4\pi} & \conj{\left(D^I_{M,K}\of{\alpha,\beta,\gamma}\right)} \op{R}\of{\alpha,\beta,\gamma} \notag \\
    & \sin\of{\beta}\wrt{\gamma}\wrt{\beta}\wrt{\alpha},
\end{align}
where $\op{R}\of{\alpha,\beta,\gamma} = R_z\of{\alpha}R_y\of{\beta}R_z\of{\gamma}$ is the rotation operator with Euler angles $\alpha,\beta,\gamma$ and $D^I_{M,K}\of{\alpha,\beta,\gamma}$ are the Wigner D-matrices.
The effect of this projection operator on an angular momentum eigenstate $\ket{I',K'}$, with quantum numbers fulfilling $\op{J}^2\ket{I',K'} = \hbar^2 I' (I'+1)\ket{I',K'}$ and $\op{J}_z\ket{I',K'} = \hbar K'\ket{I',K'}$, is then $\op{P}^I_{M,K}\ket{I',K'} = \delta_{I,I'} \delta_{K,K'}\ket{I,M}$.
In a similar way, particle number conservation is restored using the operator
\begin{equation}
    \op{P}^N = \frac{1}{2\pi}\int_0^{2\pi} \expp{i \phi \left(\op{N} - N\right)}\wrt{\phi},
\end{equation}
where $N$ is the particle number the state is projected to,
$\op{N}$ is the particle number operator, and $\phi$ is the corresponding gauge angle.
The total projection operator to a given angular momentum $I$,
angular momentum projection $M$,
as well as neutron and proton numbers $N$ and $Z$,
is then
\begin{equation}
    \op{P}^I_{M\,K} \op{P}^N \op{P}^Z.
\end{equation}

The solution is obtained
by calculating the Hamiltonian matrix elements
between the basis states and solving the resulting eigenvalue equation.
Since the states are not necessarily orthogonal,
the eigenvalue problem becomes the Hill--Wheeler equation,
\begin{equation}
    \mat{H} \vec{h}_i = E_i \mat{O} \vec{h}_i,
\end{equation}
where $\vec{h}_i$ are the Hill--Wheeler coefficients, $\mat{H}$ and $\mat{O}$ are the Hamiltonian and overlap matrices of the projected basis states $\left\{\op{P}\ket{\phi\of{\vec{q}_i}}\right\}$, defined as $H_{i j} = \braopket{\phi\of{\vec{q}_i}}{\hconj{\op{P}}\op{H}\op{P}}{\phi\of{\vec{q}_j}}$ and $O_{i j} = \braopket{\phi\of{\vec{q}_i}}{\op{P}}{\phi\of{\vec{q}_j}}$, using $\op{P}$ to represent all applied projections. The resulting energy eigenstates are then
\begin{equation}\label{eq:hwsol}
    \ket{\Psi^{I,M}_i} = \sum_{j,K} h^{I,M}_{i j,K} \op{P}^I_{M K} \op{P}^N \op{P}^Z\ket{\phi\of{\vec{q}_j}},
\end{equation}
with energies $E^I_i$ and Hill--Wheeler coefficients $h^{I,M}_{i j,K}$.
Since $\op{H}$ is rotationally invariant,
the solution is degenerate in $M$,
and so $h^{I,M}_{i j,K} = h^{I,M'}_{i j,K} \equiv h^I_{i j,K}$.
To lighten the notation, a state with unspecified $M$ is denoted $\ket{\Psi^I_i}$.

The set of basis states is constructed
by finding the HFB state that minimizes the energy
with the collective parameters as a varying constraint,
$\op{H}\of{\vec{q}} = \op{H} - \sum_i \lambda_i \op{Q}_i$,
where $\op{H}$ is the Hamiltonian of the system,
$\op{Q}_i$ are the operators defining the constraints $\expect{\op{Q}_i} = q_i$
with corresponding Lagrange multipliers $\lambda_i$.
This gives
\begin{equation}
    \braopket{\phi\of{\vec{q}_i}}{\op{H}\of{\vec{q}_i}}{\phi\of{\vec{q}_i}} = E\of{\vec{q}_i}.
\end{equation}
A term called cranking, proportional to the angular momentum around the $x$ axis $\op{j}_x$, is also added to the Hamiltonian.
This is given by $-\omega \op{j}_x$,
where $\omega$ is called the cranking frequency.
The cranking term is the intrinsic potential felt by the nucleons
(from centrifugal and Coriolis forces)
when the nucleus is cranked around the $x$ axis with frequency $\omega$.

In the method used in this article, triaxial quadrupole deformation, cranking, and proton and neutron pairing strengths are used as generator coordinates,
generating a five dimensional GCM landscape.
The deformation is introduced 
to probe deformation and triaxiality degrees of freedom $(\beta,\gamma)$,
while cranking frequency and pairing strengths are chosen randomly among three possible values for each basis state, as described in detail in \cite{Ljungberg:22,Ljungberg:24}.
These generator coordinates are chosen to capture the vibrational,
pairing vibrational, and rotational excitation modes.
The states with HFB energies within $E_\mathrm{cut}$
of the lowest one are kept,
projected to restore symmetries,
and the Hill--Wheeler equation is solved
using the projected states as a basis.
The number of kept HFB states is denoted $\ngrid$.

In order to generate the GCM basis states, an effective Hamiltonian constructed to reproduce the energy of a quadrupole constrained DFT calculation was used.
Density functionals have many useful properties,
and describe the whole nuclear chart with precision at little computational cost.
However, due to the explicit density dependence,
and in particular the presence of non--integer powers of the density,
density functionals cannot be extended
to evaluate matrix elements between different states
or projected states in a unique and straightforward way
\cite{Anguiano:01,Lacroix:09,Duguet:09,Robledo:18,Sheikh:21}.
By constructing an effective Hamiltonian from a density functional,
it inherits the wide applicability on the nuclear chart of DFT,
while still straightforward to use in the GCM.

The effective Hamiltonian used is given by
$\op{H} = \op{H}_0 + \op{H}_P + \op{H}_Q$,
consisting of a single--particle spherical reference potential $\op{H}_0$,
a pairing interaction $\op{H}_P$,
and a quadrupole--quadrupole--like interaction $\op{H}_Q$
with a modified radial dependence.
The pairing interaction used is the so called seniority pairing which is defined by
\[
    \op{H}_P = -\frac{1}{4}\sum_{ijkl} G_{ik} P_{ij} P_{kl}\hconj{a}_i \hconj{a}_j a_l a_k,
\]
where $P_{ij} = (-1)^{j_i-m_i}\delta_{(qnlj)_i,(qnlj)_j}\delta_{m_i,-m_j}$,
and $G_{ik} = G_p$ if both $i$ and $k$ correspond to proton orbitals,
while $G_{ik} = G_n$ if they correspond to neutron orbitals,
and zero otherwise.
The seniority interaction was found to couple to collective degrees of freedom such as vibrations in a similar way to realistic Argonne interaction in \cite{Idini:12,Idini:16b}.
The quadrupole--quadrupole--like interaction is given by
\[
    \op{H}_Q = -\frac{\chi}{4}\sum_{ijkl}\sum_{\mu=-2}^2 \left(Q^{2,\mu}_{ik}\conj{Q^{2,\mu}_{lj}} - Q^{2,\mu}_{il} \conj{Q^{2,\mu}_{kj}}\right)\hconj{a}_i \hconj{a}_j a_l a_k,
\]
where $\chi$ is the interaction strength and $Q^{2,\mu}_{ij}$ are
matrix elements of a quadrupole--like operator with
a modified radial dependence. For more details
on the interaction, cf. \cite{Ljungberg:22}.
The value of $\chi$ is fitted to reproduce
the energy of a quadrupole constrained DFT calculation.
The spherically symmetric HF equations are solved with the SLy4 Skyrme functional \cite{Chabanat:98} using HOSPHE \cite{Carlsson:10}
in a harmonic oscillator space with $N \le \Nmax$ shells,
where $N = 2n+l$ is the harmonic oscillator quantum number,
and the resulting $\ncut$ lowest energy states in m-scheme
are used as single--particle basis in the GCM.
The spherical reference potential $\op{H}_0$ is then given by
a sum of the HF potential of this Slater determinant
and a constant $E_0$, as
$\op{H}_0 = \sum_i \epsilon_i \hconj{a}_i a_i + E_0$,
where $\epsilon_i$ are single--particle energies
corresponding to the spherical orbital $i$.
We call this resulting Hamiltonian SLy4-H.
$\ncut$ is chosen such that the basis contains all m-values of the included j-shells,
and for simplicity, we choose $\ncut$ as close as possible to
the number of states in a harmonic oscillator and denote
the corresponding $N$ as $\Ncut$.


In order to also capture particle-hole excitations,
each HFB vacuum is also excited. The operator used to excite, which was defined in \cite{Ljungberg:22}, is
\begin{equation}
    \ket{\phi_1 (\vec q)} = \mathcal{N} e^{\hat Z} \ket{\phi_0 (\vec{q})},
    \label{eq:temp}
\end{equation}
with,
\begin{equation}
    \hat Z = \sum_{k < k'} z_{k, k'} \beta^\dagger_{k} \beta^\dagger_{k'},
\end{equation}
where $\mathcal{N}$ is a normalization constant, $z_{k,k'}$ are the weights of the quasiparticles $k$ and $k'$ which are defined in the reference basis of the vacuum $\ket{\phi_0 (\vec{q})}$.
It is of note that this operator above is the Bogoliubov Coupled Cluster (BCC) singles operator used in \cite{Signoracci:15}. In the BCC method, the weights are the unknowns to be determined variationally. Here, the choice of weights is in principle arbitrary as long as it is linearly independent. The variational solution of the Hill--Wheeler equation will provide the linear combination of basis states corresponding to specific excitations. The operator in (\ref{eq:temp}) is also corresponding to the Thouless transformation \cite{Thouless:60,Ring:80}, relating different quasi-particle vacua.
As quasiparticle weights $z_{k,k'}$ the Boltzmann distribution of quasiparticle energy excitation was chosen, with a random phase guaranteeing linear independence.
This ensures that primarily low energy quasiparticle excitations are probed.
The temperature is chosen such that the lowest energy
two-quasiparticle excitation has a given occupation
$\temp = \expp{-\Delta E/kT}$ relative to the ground state,
where $\Delta E$ is the excitation energy of the lowest two-quasiparticle excitation.
As shown in \cite{Ljungberg:22}, including temperature excitations improves the energy, convergence, and number of excitations found in the solution of the Hill--Wheeler equation.

For the odd-even case, one quasiparticle is excited
for each unprojected state from the GCM basis states \eqref{eq:temp},
and the resulting odd particle number states are used as the basis.
The quasiparticle is chosen randomly and uniformly from the
$\qnum$ lowest energy quasiparticle excitations
with the correct symmetries.
Signature is not broken by the HFB calculation,
and all even-even vacua generated have signature $+1$,
so one signature $\sig$ is chosen and only
quasiparticle excitations with that signature are considered.
If nothing else is specified, $\sig=+i$ is used in this article.

Due to the cranking, the quasiparticles with $\sig=+i$
and $\sig=-i$ generate two sets of basis states
that give rise to different solutions in independent calculations.
Since they should converge to the same solution in an exact calculation,
the difference between the two solutions can indicate a degree of convergence,
and provide a preliminary estimate of the uncertainty of the GCM many--body calculation. For a given model space $\Ncut$, the two signatures will be independent calculations that will converge for sufficiently large generator coordinate space $\ngrid$ independently from the choice of the other parameters.

\subsection{Green's functions}

In this article, the Green's function formalism is used
to construct the effective optical potential,
using information from the nuclear structure wavefunctions. The potential will then be used to calculate scattering cross sections.

For a given Hamiltonian $\op{H}$ and ground state
$\ket{\Psi_0}$ with energy $E_0$,
the matrix elements of the time--ordered Green's function
for that Hamiltonian can be written as
\begin{equation}\label{eq:greens}
    \begin{split}
        G_{\alpha,\beta}\of{E}
        = \lim_{\eta \rightarrow 0^+} & \bra{\Psi_0}{a_\alpha\frac{1}{E-\left(H-E_0\right)+i\eta}\hconj{a}_\beta}\ket{\Psi_0} + \\%
        +                             & \bra{\Psi_0}{\hconj{a}_\beta\frac{1}{E+\left(H-E_0\right)-i\eta}a_\alpha}\ket{\Psi_0},
    \end{split}
\end{equation}
where $\eta$ is an infinitesimal term that ensures the correct time--ordering.
The matrix elements then define the operator $G$
which acts on the space of single-particle states
defined by the set of operators $a_\alpha$.

The single-particle optical potential $V$ is then
expressed as the sum of a static potential,
$V_0$, and the self energy $\Sigma$.
Using the static potential the Green's function $G_0$ is defined
for the Hamiltonian $H_0 = T + V_0$ as above,
where $T$ is the kinetic part of the Hamiltonian,
while the total Hamiltonian $H = T + V_0 + \Sigma$
defines the Green's function $G$.
These Green's functions can be related using the Dyson equation,
\begin{equation}
    G = G_0 + G_0 \Sigma G,
\end{equation}
which can be solved for $\Sigma$, as
\begin{equation}\label{eq:sigma}
    \Sigma = G_0^{-1} - G^{-1}.
\end{equation}
Since $G$ is the Green's function of $H$,
it does not depend on how one chooses to
decompose the $V$ into $V_0$ and $\Sigma$.
Therefore, $V_0$ can be chosen arbitrarily,
while the sum $V_0 + \Sigma$ will remain unchanged.
For convenience and comparison, $H_0$ was chosen as the
spherical reference potential used as the
single-particle part of the effective Hamiltonian
in the GCM calculation.

The ground state of an even--even nucleus $\ket{\Psi_0}$
has spin $0$. Since the Hamiltonian is rotationally invariant,
the analysis can be restricted to a single angular momentum $J = J_\alpha = J_\beta$,
where $J_\alpha$ is the total angular momentum quantum number
corresponding to the eigenstate $\alpha$.

To calculate the two parts of the Green's function
a complete set of states of the odd system with
one more (less) particle, $\ket{{\Psi^+}^J_i}$ ($\ket{{\Psi^-}^J_i}$),
where $H \ket{{\Psi^\pm}^J_i} = {E^\pm}^J_i \ket{{\Psi^\pm}^J_i}$,
can be used to construct the identity operators $\id^\pm$,
\begin{equation}\label{eq:completely_complete}
    \id^\pm = \sum_i \ket{{\Psi^\pm}^J_i}\bra{{\Psi^\pm}^J_i},
\end{equation}
which act on the Hilbert space of $A\pm1$ particles.


With such a basis, the expression for the identity can
be inserted into the first term of \eqref{eq:greens}, giving
\begin{widetext}
\begin{align}
\bra{\Psi_0}{a_\alpha\frac{1}{E-\left(H-E_0\right)+i\eta} \left(\sum_i \ket{{\Psi^+}^J_i}\bra{{\Psi^+}^J_i}\right)\hconj{a}_\beta}\ket{\Psi_0} = &
     \sum_i \frac{\braopket{\Psi_0}{a_\alpha}{{\Psi^+}^J_i}\braopket{{\Psi^+}^J_i}{\hconj{a}_\beta}{\Psi_0}}{E-\left({E^+}^J_i-E_0\right)+i\eta} = \notag \\
= &    \sum_i \frac{\conj{\sigma^+}_{J,i,\alpha}\sigma^+_{J,i,\beta}}{E-\left({E^+}^J_i-E_0\right)+i\eta}
 \label{eq:greensum}
\end{align}
\end{widetext}
where $\sigma^+_{J,i,\alpha}\equiv\braopket{{\Psi^+}^J_i}{\hconj{a}_\alpha}{\Psi_0}$,
(while $\sigma^-_{J,i,\alpha}\equiv\braopket{\Psi_0}{\hconj{a}_\alpha}{{\Psi^-}^J_i}$)
are the spectroscopic amplitudes,
and $\left|\sigma^\pm_{J,i,\alpha}\right|^2$ are called spectroscopic factors.
The second term of \eqref{eq:greens} can,
in the same way, be written as
\[
    \sum_i \frac{\conj{\sigma^-}_{J,i,\alpha}\sigma^-_{J,i,\beta}}{E+\left({E^-}^J_i-E_0\right)-i\eta}.
\]


Defining
$(\bar{E}^\pm)^{J\pi}_{i} \of{\eta} = \pm\left({E^\pm}^{J\pi}_{i} - E_0 - i\eta\right)$,
Eq. \eqref{eq:greens} can then be written as
\begin{equation}\label{eq:finalgreens}
    G^{J\pi}_{\alpha,\beta}\of{E} = \lim_{\eta \rightarrow 0^+}\sum_{i,s=\pm} \frac{\conj{\sigma^s}_{J,i,\alpha}\sigma^s_{J,i,\beta}}{E-(\bar{E}^s)^{J\pi}_{i}\of{\eta}},
\end{equation}
where the sum of $s$ over $+$ and $-$ denotes a sum over $A+1$ and $A-1$ intermediate states respectively.

In the continuum, the above sums becomes integrals over scattering states, and this together with the infinitesimal $\eta$ gives a non--Hermitian contribution to the Green's function. This contribution describes the coupling between the basis states and the continuum. However, a calculation in a discrete basis necessarily produces a discrete set of solutions.
In this article, the effect of the continuum solutions are approximated by using a finite value of $\eta$. This is equivalent to broadening each discrete state according to a Breit--Wigner shape with widths proportional to $\eta$, effectively treating them as resonances with constant or parameterized widths. This is further discussed in Appendix \ref{sec:imag_part}. It is possible to extend the method to calculate the resonance widths microscopically, however in Appendix \ref{sec:imag_part} it is shown that the impact of a different $\eta$ is not the main source of uncertainty.
$\eta$ was chosen as
\begin{equation}\label{eq:eta}
    \eta=\frac{a}{\pi}\frac{(E-E_F)^2}{(E-E_F)^2 + b_\eta^2},
\end{equation}
where $E$ is the energy of the scattering particle, $E_F$ is the Fermi energy of the target nucleus taken as the negative average of the calculated neutron separation energies in $\mg{24,25}$, and $a$ and $b_\eta$ are parameters. The parameters were chosen as $a=\MeV{12}$ and $b_\eta=\MeV{22.36}$ as in \cite{Waldecker:11,Brown:81} for all of the results, and the dependence on $\eta$ was investigated by varying $a$.

To use this formalism of Eq. \eqref{eq:finalgreens} to
calculate Green's functions
using a nuclear many--body method,
the method needs to provide the energy levels and
their spectroscopic amplitudes relative to
the state on which the particle is scattering.

\subsection{Calculation of the Spectroscopic amplitudes}

Calculating spectroscopic amplitudes in a multi-reference basis
requires calculating matrix elements
between all pairs of projected basis states.
It is therefore important to exploit the properties of
the projection and overlap in order to
decrease the computational complexity as much as possible.
To do this, the formula for the overlap \cite{Carlsson:21}
can be rewritten to efficiently calculate
spectroscopic amplitudes between a given pair of basis states.

Given the energy eigenstates
found by solving the Hill-Wheeler equation,
the spin 0 $A$--particle ground state
$\ket{\Psi^{0,0}_0} = \sum_{a'} h^0_{0 a',0} \op{P}^0_{0 0} \op{P}^A\ket{\phi_{a'}}$ and the $A+1$--particle state $\ket{{\Psi^+}^{I,M}_i} = \sum_{a,K} h^I_{i a,K} \op{P}^I_{M K} \op{P}^{A+1} \hconj{\beta}_\gamma \ket{\phi_a}$, where $\op{P}^A = \op{P}^N \op{P}^Z$ and $\op{P}^{A+1} = \op{P}^{N+1} \op{P}^Z$ or $\op{P}^N \op{P}^{Z+1}$ depending on whether $\hconj{a}_\alpha$ creates a neutron or proton,
and the excited quasiparticle $\hconj{\beta}_\gamma$ is chosen based on each $\ket{\phi_a}$, the spectroscopic amplitudes can be calculated as
\begin{align*}
 \sigma^+_{J,i,\alpha} = \braopket{{\Psi^+}^{I,M}_i}{\hconj{a}_\alpha}{\Psi^{0,0}_0} = \notag \\
 = \sum_{a,a',K} \left(h^I_{i a,K}\right)^* h^0_{0 a',0} 
    \braopket{\phi_a}{\beta_\gamma \op{P}^{A+1} \op{P}^I_{K M}\hconj{a}_\alpha \op{P}^0_{0 0} \op{P}^A}{\phi_{a'}}.
\end{align*}
Creating a particle in the state $\alpha$ on a spin 0 state, leaves the state with spin $j_\alpha$ and spin projection $m_\alpha$. The projection operator $\op{P}^I_{K M}$ will then project this to spin projection $K$, giving $\op{P}^I_{K M} \hconj{a}_\alpha \op{P}^0_{0 0} = \hconj{a}_{\alpha_K} \op{P}^0_{0 0} \delta_{I,j_\alpha} \delta_{M,m_\alpha}$, where $\hconj{a}_{\alpha_K}$ is the creation operator with the same quantum numbers as $\alpha$ except $K$ substituted for $m_\alpha$.
Additionally, $\op{P}^{A+1} \hconj{a}_\alpha \op{P}^A = \hconj{a}_\alpha \op{P}^A$, since $\hconj{a}_\alpha \op{P}^A$ always creates a state with $A+1$ particles.
With these relations, the terms of the sum can be simplified to
\begin{align*}
& \braopket{\phi_a}{\beta_\gamma \op{P}^{A+1} \op{P}^I_{K M}\hconj{a}_\alpha \op{P}^0_{0 0} \op{P}^A}{\phi_{a'}} = \notag \\
= & \braopket{\phi_a}{\beta_\gamma \hconj{a}_{\alpha_K} \op{P}^0_{0 0} \op{P}^A}{\phi_{a'}}.
\end{align*}
Then, $\hconj{a}_\alpha$ can be expressed in terms of the quasiparticle creation and annihilation operators of $\ket{\phi_a}$ using the Bogoliubov transformation
\[
    \hconj{a}_\alpha = \sum_\delta {U_{\alpha \delta}}^* \hconj{\beta}_\delta + V_{\alpha \delta} \beta_\delta,
\]
where $U$ and $V$ are the Bogoliubov matrices defining the state $\ket{\phi_a}$.
$\bra{\phi_a}\beta_\gamma \hconj{a}_\alpha$ can therefore be expanded into a linear combinations of the terms $\bra{\phi_a}\beta_\gamma \hconj{\beta}_\delta = \delta_{\gamma \delta}\bra{\phi_a}$ and $\bra{\phi_a}\beta_\gamma \beta_\delta$. \cite{Bostrom:23proc}

In order to calculate matrix elements of the form $\braopket{\phi}{\beta_\gamma \beta_\delta}{\phi'}$ efficiently, the method outlined in \cite{Bostrom:23proc} was used. The Bloch--Messiah decompositions of the Bogoliubov transformation describing $\ket{\phi}$ and $\ket{\phi'}$ are defined to be $(U=D\bar{U}C,V=\conj{D}\bar{V}C)$ and $(U'=D'\bar{U}'C',V'=\conj{D'}\bar{V}'C')$, respectively. The quasiparticle operators are then written as
\begin{equation}
    \hconj{\beta}_\gamma = \sum_i \mathbf{U}^\gamma_i\hconj{a}_i + \mathbf{V}^\gamma_i a_i \quad \text{and} \quad \hconj{\beta}_\delta = \sum_i \mathbf{U}^\delta_i\hconj{a}_i + \mathbf{V}^\delta_i a_i,
\end{equation}
where $\mathbf{U}^\gamma$, $\mathbf{V}^\gamma$, $\mathbf{U}^\delta$, and $\mathbf{V}^\delta$
are vectors defining the quasiparticle operators $\beta_\gamma$ and $\beta_\delta$,
and are in this case columns of the matrices $U$ and $V$ defining $\ket{\phi}$.
$\braopket{\phi}{\beta_\gamma \beta_\delta}{\phi'}$ can then be calculated as
\begin{equation}\label{eq:qp_pf}
    \braopket{\phi}{\beta_\gamma \beta_\delta}{\phi'} = (-1)^{n/2}\pf\begin{pmatrix}
        \mathcal{A}      & \mathcal{B} \\
        -\T{\mathcal{B}} & \mathcal{C}
    \end{pmatrix}
\end{equation}
where $\pf$ denotes the Pfaffian. The matrix $\mathcal{A}$,
\begin{equation}
    \mathcal{A}=-\T{\mathcal{A}}=\begin{pmatrix}
        -[\bar{U}\sigma]_{n\times n} & [\Lambda \hconj{D} D' \Lambda']_{n\times n'} \\
                                     & [\sigma\bar{U}']_{n'\times n'}
    \end{pmatrix}
\end{equation}
only depends on $\ket{\phi}$ and $\ket{\phi'}$, and
\begin{align}
    \mathcal{B}& =\begin{pmatrix}
        [\Lambda\hconj{D}{\mathbf{V}^\gamma}^*]_{n\times 1} & [\Lambda\hconj{D}{\mathbf{V}^\delta}^*]_{n\times 1} \\
        -[\Lambda'\T{D'}{\mathbf{U}^\gamma}^*]_{n\times 1}  & -[\Lambda'\T{D'}{\mathbf{U}^\delta}^*]_{n\times 1}
    \end{pmatrix},
\notag \\
    \mathcal{C}& =-\T{\mathcal{C}}=\begin{pmatrix}
        0 & [\hconj{\mathbf{U}^\gamma}{\mathbf{V}^\delta}^*]_{1\times 1} \\
          & 0
    \end{pmatrix}
\end{align}
contain the contribution from the quasiparticles. Here $\Lambda$ is a diagonal matrix with the diagonal elements $\sqrt{v_0},\sqrt{v_0},\sqrt{v_1},\sqrt{v_1},\dots$, with $v_i$ being the elements of the $\bar{V}$, and $\sigma$ is a block diagonal matrix with each block equal to $\begin{pmatrix}
    0  & 1 \\
    -1 & 0
\end{pmatrix}$.
The notation $[\cdot]_{n\times n'}$ denotes truncating
the matrix to size $n\times n'$ as explained in \cite{Carlsson:21},
and is possible since unoccupied single-particle states
do not contribute the overlap.
The Pfaffian in \eqref{eq:qp_pf} can then be computed using the identity
\begin{equation}\label{eq:pf_sep}
    \pf\begin{pmatrix}
        \mathcal{A}      & \mathcal{B} \\
        -\T{\mathcal{B}} & \mathcal{C}
    \end{pmatrix} = \pf\of{\mathcal{A}}\pf\of{\mathcal{C} + \T{\mathcal{B}}\inv{\mathcal{A}}\mathcal{B}}
\end{equation}
where $\mathcal{C} + \T{\mathcal{B}}\inv{\mathcal{A}}\mathcal{B}$
is a 2 by 2 matrix, and its Pfaffian is simply given by
its top right matrix element.
Since $\mathcal{A}$ does not depend on
the quasiparticle operators $\beta_\gamma \beta_\delta$,
\eqref{eq:pf_sep} makes it possible to calculate
$\pf\of{\mathcal{A}}$ and $\inv{\mathcal{A}}$ once
and then calculate the top right element of
$\mathcal{C} + \T{\mathcal{B}}\inv{\mathcal{A}}\mathcal{B}$
for each quasiparticle pair.
This procedure makes it practically possible to compute the large
number of spectroscopic amplitudes required in this framework.

\subsection{Sum rules and completion}

Using a many--body method, such as the GCM procedure described above,
one can find the ground state of the $A$-particle nucleus,
the spectra of both $(A+1)$- and $(A-1)$-particle nuclei,
as well as the spectroscopic amplitudes $\sigma^\pm_{J,i,\alpha}$.
However, one might not get a complete basis to fulfill
\eqref{eq:completely_complete} and be able to construct the identity
on the space of $A\pm 1$ wavefunctions.
In general, the GCM procedure used in this article does not
provide a complete basis,
since states with $A$ and $A+1$ particles are obtained separately,
and the full $\Nmax$ many--body space would need to be spanned to achieve completeness.
But to derive the Lehmann representation \eqref{eq:greensum} from the general form \eqref{eq:greens} only requires that
\begin{equation}\label{eq:almost_completeness}
    \begin{split}
        \sum_i \ket{{\Psi^+}^J_i}\braopket{{\Psi^+}^J_i}{\hconj{a}_\alpha}{\Psi_0} & = \hconj{a}_\alpha\ket{\Psi_0}, \\
        \sum_i \ket{{\Psi^-}^J_i}\braopket{{\Psi^-}^J_i}{a_\alpha}{\Psi_0}         & = a_\alpha\ket{\Psi_0}.
    \end{split}
\end{equation}
That is, the basis only needs to be complete on the smaller space of
single--particle excitations of the state $\ket{\Psi_0}$,
since the Green's function only describe the propagation of single--particle excitations.


In the case where \eqref{eq:almost_completeness} holds, the following well known sum rule can be derived,
\begin{align}\label{eq:test}
        \delta_{\alpha,\beta} = & \braopket{\Psi_0}{\acomm{a_\alpha}{\hconj{a}_\beta}}{\Psi_0} = \notag \\ 
        & \sum_{i} \braopket{\Psi_0}{a_\alpha}{{\Psi^+}^J_i}\braopket{{\Psi^+}^J_i}{\hconj{a}_\beta}{\Psi_0} \notag \\+&
        \sum_{i} \braopket{\Psi_0}{\hconj{a}_\beta}{{\Psi^-}^J_i}\braopket{{\Psi^-}^J_i}{a_\alpha}{\Psi_0} \notag \\ =&
        \sum_{i,s=\pm} \conj{\sigma^s}_{J,i,\alpha}\sigma^s_{J,i,\beta}.
\end{align}
Furthermore, one can consider the energy--weighted sum--rule introduced in 
\cite{Bostrom:25},
\begin{widetext}
\begin{equation}\label{eq:energy_sumrule}
    \braopket{\Psi_0}{\acomm{a_\alpha}{\comm{\op{H}}{\hconj{a}_\beta}}}{\Psi_0} = \sum_i \conj{\sigma^+}_{J,i,\alpha} ({E^+}^J_i - E_0) \sigma^+_{J,i,\beta} - \sum_i \conj{\sigma^-}_{J,i,\alpha} ({E^-}^J_i - E_0) \sigma^-_{J,i,\beta},
\end{equation}
\end{widetext}
relating the spectroscopic amplitudes and energy of the states with the Hamiltonian properties. Here, $\left(H_A\right)_{\alpha,\beta} \equiv \braopket{\Psi_0}{\acomm{a_\alpha}{\comm{\op{H}}{\hconj{a}_\beta}}}{\Psi_0}$ can be seen as the matrix elements of a generalized mean--field for the state $\ket{\Psi_0}$ and its eigenvalues can be experimentally observed as the energy centroids for the total stripping and pick--up strengths of the single--particle states \cite{Rowe:68}.
$H_A$ will be called the many--body mean--field in the rest of the article. The effect on the cross section of possible approximations to $H_A$ are discussed in Appendix \ref{sec:completion}.

These sum rules also correspond to
the first and second terms in the asymptotic
expansion of the Green's function,
which can be seen by looking at
the asymptotic expansion of the sum \eqref{eq:greensum}
\begin{align}\label{eq:sumrule_asymptotic}
    \sum_i \frac{\conj{\sigma^+}_{J,i,\alpha}\sigma^+_{J,i,\beta}}{E-\left(E^J_i-E_0\right)\pm i\eta} \approx \frac{1}{E}\sum_i \conj{\sigma^+}_{J,i,\alpha}\sigma^+_{J,i,\beta} + \notag \\ 
    \frac{1}{E^2}\sum_i \conj{\sigma^+}_{J,i,\alpha}\left(E^J_i-E_0\right)\sigma^+_{J,i,\beta} + \mathcal{O}\of{\frac{1}{E^3}},
\end{align}
assuming $\eta \ll E$ and $\eta \ll E^J_i-E_0$.
The first term is exactly \eqref{eq:test} divided by $E$,
and the second term gives
the energy--weighted sum rule divided by $E^2$.

If the sum rules (\ref{eq:test},\ref{eq:energy_sumrule}) do not hold,
the calculation is missing an unknown number of relevant states,
$\ket{\Phi_i}$, with corresponding energies ${E_\Phi}_i$,
which fulfill \eqref{eq:almost_completeness} together with the approximate solutions found $\ket{\Psi^J_i}$.
In \cite{Bostrom:25}, the contribution of the missing states was proposed to be approximated by a minimal set of states, interpreted as a mean--field average of all missing states.
In fact, due to the asymptotic expression \eqref{eq:sumrule_asymptotic},
this average contribution will give the correct contribution to the Green's function up to $\mathcal{O}\of{\Delta E^{-3}}$, where $\Delta E$ is the energy difference between the state and the energy of the scattering particle, and so for states far away from the scattering energy the error approaches zero.


To find states which complete the sum rules,
which are hereafter called completion states,
the sum rules \eqref{eq:test} and \eqref{eq:energy_sumrule}
are expressed in matrix form using
$\sigma_{i,\alpha}=\sigma^{s_i}_{J,i,\alpha}$ and
$\epsilon_{i,j} = s_i (E^J_i - E_0)\delta_{i,j}$
where $s_i=+$ for particle states and $s_i=-$ for hole states.
Additionally, the spectroscopic amplitudes of the completion states
in matrix form are denoted as
$c_{i,\alpha} = \braopket{\Phi_i}{\hconj{a}_\alpha}{\Psi_0}$,
and their energies as
$\left(\epsilon_c\right)_{i,j} = {E_\Phi}_i\delta_{i,j}$.
Equations \eqref{eq:test} and \eqref{eq:energy_sumrule} then become
\begin{equation}\label{eq:complete_matrixeq}
    \hconj{\sigma}\sigma + \hconj{c}c = \id{}
\end{equation}
and
\begin{equation}\label{eq:complete_energy_sumrule}
    \hconj{\sigma} \epsilon \sigma + \hconj{c} \epsilon_c c = H_A.
\end{equation}

$H_A$ can be obtained by explicitly calculating $\braopket{\Psi_0}{\acomm{a_\alpha}{\comm{\op{H}}{\hconj{a}_\beta}}}{\Psi_0}$ for the Hill--Wheeler ground state.
For the one--body term $\op{\Gamma} = \sum_{\alpha \beta} \Gamma_{\alpha \beta} \hconj{a}_\alpha a_\beta$, the commutator can be calculated to give
$\braopket{\Psi_0}{\acomm{a_\alpha}{\comm{\op{\Gamma}}{\hconj{a}_\beta}}}{\Psi_0} = \Gamma_{\alpha \beta}$,
where $\Gamma_{\alpha \beta}$ are the matrix elements of $\op{\Gamma}$ in the single--particle basis defined by $a_\alpha$.
For the two--body terms, the commutator expression can be expanded
into a one--body operator,
and calculated using the same projection procedure
used to calculate the Hamiltonian and overlap as described above.
To find this one--body operator for
a general antisymmetrized two--body interaction described by
$\op{V} = \frac{1}{4}\sum_{i j k l}\bar{v}_{i j k l} \hconj{a}_i \hconj{a}_j a_l a_k$,
the commutator can explicitly be calculated as
\begin{align}
    \frac{1}{4}\sum_{i j k l}\bar{v}_{i j k l}\acomm{a_\alpha}{\comm{\hconj{a}_i \hconj{a}_j a_l a_k}{\hconj{a}_\beta}} = \notag \\
   = \frac{1}{4}\sum_{i j l}\bar{v}_{i j k l}\acomm{a_\alpha}{\delta_{k,\beta}\hconj{a}_i\hconj{a}_j a_l - \delta_{l,\beta}\hconj{a}_i\hconj{a}_j a_k},
\end{align}
using the fermionic anticommutation relations for $\hconj{a}$ and $a$,
where the right hand side is equal to
$\frac{1}{2}\sum_{i j l}\bar{v}_{i j \beta l}\acomm{a_\alpha}{\hconj{a}_i\hconj{a}_j a_l}$,
due to the antisymmetry of $\bar{v}_{i j k l} = -\bar{v}_{i j l k} = -\bar{v}_{j i k l}$.
The anticommutator can then in turn be expanded as
\begin{align}
    \frac{1}{2}\sum_{i j l}\bar{v}_{i j \beta l}\acomm{a_\alpha}{\hconj{a}_i\hconj{a}_j a_l} = \notag \\
  = \frac{1}{2}\sum_{i j l}\bar{v}_{i j \beta l}\left(\delta_{i,\alpha}\hconj{a}_j a_l - \delta_{j,\alpha}\hconj{a}_i a_l\right),
\end{align}
where, again, one can exploit the antisymmetry of $\bar{v}$ to finally get
\begin{equation}
    \acomm{a_\alpha}{\comm{\op{V}}{\hconj{a}_\beta}} = \sum_{i j}\bar{v}_{\alpha i \beta j}\hconj{a}_i a_j.
\end{equation}
In this form it becomes clear that $H_A$ is the average mean-field corresponding to the full many-body wavefunction 
\cite{Rowe:68}.




Completing with the same number of states as
shells of the specific angular momentum $J$,
so that $c$ is a square matrix,
\eqref{eq:complete_matrixeq} together with \eqref{eq:complete_energy_sumrule}
determine $c$ and $\epsilon_c$ completely.
To see this, one can look at the polar decomposition of $c = U P$,
where $U$ is unitary and $P$ is Hermitian and positive (semi-)definite.
Then, \eqref{eq:complete_matrixeq} gives
\[
    P^2 = \id{} - \hconj{\sigma}\sigma.
\]
Since the eigenvalues of $\hconj{\sigma}\sigma$ are occupation probabilities, they are guaranteed to be between 0 and 1.
$P$ can therefore be calculated as the matrix square root of $\id{}-\hconj{\sigma}\sigma$ which is Hermitian and positive semidefinite.
Inserting this expression for $P$ into
\eqref{eq:complete_energy_sumrule} then gives
\[
    \hconj{U} \epsilon_c U = P^{-1} \left(\epsilon_0 - \hconj{\sigma} \epsilon \sigma\right) P^{-1}.
\]
Since the right hand side is Hermitian, $\epsilon_c$ is diagonal, and $U$ is unitary,
the left hand side is exactly the diagonalization of the right hand side.
Since the diagonalization is unique, up to permutations and degeneracy,
this completely determines both the spectroscopic amplitudes $c$ and
the energies of the completion levels up to reordering the levels,
which does not affect any observable.

This choice of $c$ can be further understood by considering the case of a single--particle Hamiltonian.
In that case, the only states with nonzero spectroscopic factors are single--particle excitations,
and finding the completion spectroscopic amplitudes and energies from the sum rules \eqref{eq:test} and \eqref{eq:energy_sumrule} reduce to diagonalizing the single--particle Hamiltonian,
where now the spectroscopic amplitudes is simply the coefficients of the single--particle wavefunctions,
and $\braopket{\Psi_0}{\acomm{a_\alpha}{\comm{\op{H}}{\hconj{a}_\beta}}}{\Psi_0}$ are the matrix elements of the single--particle Hamiltonian.

    \begin{figure*}[ht]%
        \centering%
        \includegraphics[width=0.7\textwidth]{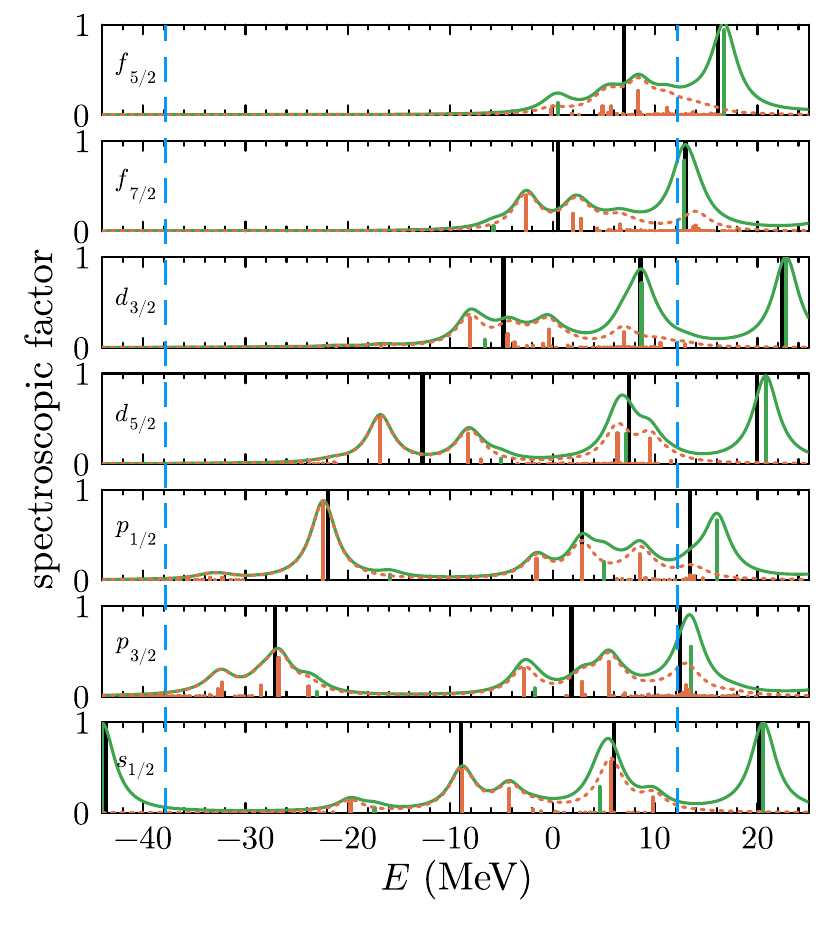}%
        \caption{ 
    Comparison of spectroscopic strength as a function of energy for different calculations. i) Neutron spectroscopic factors for each GCM state $i$ of $\mg{25}$ plotted at $E=E^+_i-E_0$, appear above the Fermi energy $E_F=\MeV{-12.78}$, and each $\mg{23}$ state $j$ plotted at $E=-(E^-_j-E_0)$ appear below, shown as orange bars; ii) the completion states shown as green bars; iii) single--particle states of $\mg{24}$ for the many-body mean-field $H_A$ shown with black bars.
    The lines are obtained by convolution with a Lorentzian, $\frac{\eta^2}{E^2+\eta^2}$
    with $\eta=\MeV{1.5}$, to approximate the resonance widths.
    This gives the dashed orange line for only the GCM states,
    and the green solid line when also including the completion states.
    The levels are grouped by total orbital angular momentum and total spin.
    An energy range of $\MeV{50}$ is shown around the Fermi energy with two dashed blue vertical lines, corresponding the energy cutoff of $\MeV{25}$ used in the selection of basis states used in the GCM calculation.
}%
        \label{fig:biggrid_sf2_combined_edited_nog}%
    \end{figure*}

\autoref{fig:biggrid_sf2_combined_edited_nog} shows the result of a calculation and its completion states. The height of each bar correspond to the spectroscopic factor of each state. The spectroscopic factors obtained from the GCM are presented in orange while the completion is in green. The spectroscopic factors are also shown convolved with a Lorentzian, $\frac{\eta}{\pi} \frac{1}{E^2+\eta^2}$, demonstrating the effect of a finite $\eta$. $\eta$ was chosen as $\MeV{1.5}$, which corresponds to an energy around $\MeV{18}$ above the neutron Fermi energy, or around $\MeV{5}$ of neutron scattering energy, when using Eq. \eqref{eq:eta}. These are compared to the energy levels obtained by diagonalizing the many-body mean-field $H_A$, where all spectroscopic factors are by definition equal to one. One can see that for a large energy range, the GCM states fulfill the sum rule well and the completion levels correspondingly have small spectroscopic factors, but at energies further from the Fermi energy, the completion is required to fulfill the sum rules, and the energies of the completion states are close to the single-particle energies.

This procedure enables using low energy many body information to describe single--particle processes, and in particular the scattering of neutrons, by complementing the many body method with information about the mean--field single--particle potential.

\subsection{Regularization}

Reconstructing the real-space or k-space potentials from their operator
on a truncated space is not straightforward.
In principle, the potential operator can be expanded in a basis as
\begin{equation}
    \op{V} = \sum_{n n'} V_{n n'}\ket{n}\bra{n'}
\end{equation}
where $V_{n n'} = \braopket{n}{\op{V}}{n'}$,
and from this the non-local momentum space potential can be written as
\begin{align}\label{eq:basis2spatial}
    V\of{k, k'} & =
    \braopket{k}{\op{V}}{k'} =
    \sum_{n n'} V_{n n'} \braket{k}{n} \braket{n'}{k'} = \notag \\
    & = \sum_{n n'} \psi_n\of{k} V_{n n'} \conj{\psi_{n'}\of{k'}}.
\end{align}
This converges formally when the size of the basis increases,
but truncating this sum at $N_{max}$ introduces artifacts.
This occurs in a similar fashion as the Gibbs phenomenon in
Fourier series, and appears as large amplitude oscillations in the potential.
While not a problem when calculating in the truncated basis,
the oscillations will introduce unphysical effects
if the optical potential in momentum or real space representation
would be used directly in a scattering calculation,
since they cause strong couplings to the scattering states,
leading to effects on the cross sections.

The problem can be mitigated by smoothing out the potential
by multiplying with a regulator such as
\begin{equation}\label{eq:smooth_sigma}
    \sigma_n = \frac{1 - \exp\of{-\left(\alpha \frac{n - N - 1}{N}\right)^2}}{1 - \exp\of{-\alpha^2}},
\end{equation}
where $\alpha$ is a dimensionless smoothing parameter, usually chosen between 5 and 10,
so that the smoothed potential becomes
\begin{equation}
    \tilde{V}_{n_1\,n_2} \rightarrow \sigma_{n_1} V_{n_1\,n_2} \sigma_{n_2}.
\end{equation}
Since $\sigma_n \rightarrow 1$ as $N\rightarrow\infty$,
this still converges to the correct potential when the basis size
is increased, but it dampens the high-frequency components that give
rise to the oscillations \cite{Revai:85,shirokov:2021,du:2022}.

\subsection{Scattering}

With the static potential $V_0$ together with the self energy
$\Sigma$, the total optical potential $V = V_0 + \Sigma$
is then used to calculate scattering cross-sections.
This is done by evaluating the optical potential in k-space
for each partial wave $l,j$,
and rewriting the Lippmann-Schwinger equation:
\begin{equation}\label{eq:lippmann_schwinger}
    T_{l,j} = V_{l,j} + V_{l,j} G_0 T_{l,j},
\end{equation}
where $G_0$ is the free particle Green's function
and $T_{l,j}$ is the transmission matrix for the given partial wave,
as
\begin{equation}\label{eq:lippmann_schwinger_inv}
    T_{l,j} = \left(\id - V_{l,j} G_0\right)^{-1} V_{l,j}.
\end{equation}
The phase shifts can then be calculated as
\begin{widetext}
\begin{equation}
    \expp{2 i \delta_{l,j}} = \frac{g'_l\of{R_C k_0}\left(S_{l,j}\of{k_0,k_0}-1\right) + i j'_l\of{R_C k_0}\left(S_{l,j}\of{k_0,k_0}+1\right)}{g_l\of{R_C k_0}\left(S_{l,j}\of{k_0,k_0}-1\right) + i j_l\of{R_C k_0}\left(S_{l,j}\of{k_0,k_0}+1\right)},
\end{equation}
\end{widetext}
where $S_{l,j}\of{k_0,k_0} = 1 - 2i\pi T_{l,j}\of{k_0,k_0}$
is the scattering S-matrix,
with $g_l(R_C k_0)$ and $j_l(R_C k_0)$ are Bessel functions of the first and second kind respectively, and their derivatives $g'_l$, $j'_l$, all calculated at a matching radius $R_C$. Results are independent of matching radius as long as the asymptotic region is reached. Here we use a large $R_C = 18.6\unit{fm}$ throughout
\cite{Descouvemont:10}.

Following \cite{Arellano:21}, the differential cross section of the neutron can then be calculated as
\[
    \frac{\mathrm{d}\sigma}{\mathrm{d}\Omega} = \left|g\of{\theta}\right|^2+\left|h\of{\theta}\right|^2,
\]
where $g\of{\theta}$ is given by
\begin{equation}
    g\of{\theta} = \frac{i}{4k}\sum_{l=0}\sum_j (2j+1)\left(1-\expp{2i \delta_{j l}}\right) P_l\of{\cos\of{\theta}},
\end{equation}
and $h\of{\theta}$ is given by
\begin{equation}
    h\of{\theta} = -\frac{1}{2k}\sum_{l=1}\sum_j C_{j l}\left(1-\expp{2i \delta_{j l}}\right) \frac{\wrt{P_l\of{\cos\of{\theta}}}}{\wrt{\theta}},
\end{equation}
where
\begin{equation}
    C_{j l} = \begin{cases}
        -1\mathrm{\qquad for \quad } j = l - \sfrac{1}{2} \\
        +1\mathrm{\qquad for \quad } j = l + \sfrac{1}{2} \\
    \end{cases}.
\end{equation}
Integrated cross sections can also be calculated from the phase shifts as
\begin{align}
    \sigma_R & = \frac{\pi}{2 k^2}\sum_{l=0}\sum_j (2j+1)\left(1-\left|\expp{2 i \delta_{j l}}\right|^2\right) \\
    \sigma_E & = \frac{\pi}{2 k^2}\sum_{l=0}\sum_j (2j+1)\left|1-\expp{2 i \delta_{j l}}\right|^2              \\
    \sigma_T & = \frac{\pi}{k^2}\sum_{l=0}\sum_j (2j+1)\left(1-\Re\of{\expp{2 i \delta_{j l}}}\right),
\end{align}
where $\sigma_R$ is the reaction cross section,
$\sigma_E$ is the elastic cross section,
and $\sigma_T$ is the total cross section.

We employ the center of mass correction described in Eq. (8) in \cite{Bostrom:25},
and all the scattering energies are given in the laboratory frame in this article.

\section{Neutron scattering on $\isotope{48,50}{Cr}$}
\label{sec:Cr}

Chromium bearing steel alloys with 10--16\% Cr content are widely used as reactor casing and other structural materials in fission and fusion devices for their suitable thermal properties. For neutron--induced reactions, the energy region roughly between 1 and 15~MeV is particularly relevant, because it covers most of the fission and fusion fast neutron spectrum (with $d{+}t$ fusion producing a prominent 14.1~MeV peak). Recently, the impact of chromium cross section on the effective critical mass of a model reactor was found non negligible, with differences among nuclear data libraries yielding up to 10\%  difference in effective critical mass \cite{Koscheev:17}.
Experimental determinations of cross sections and angular distributions in this range are often limited by the preparation of the target (e.g. oxidation and target thickness), which propagate into both the absolute normalization and the shape of differential cross sections. In steel benchmarks, uncertainties in $^{56}$Fe and chromium cross sections can partially compensate each other. As the $^{56}$Fe evaluations have been tightened (e.g., by the CIELO effort \cite{Herman:18}), the chromium isotopes neutron cross sections now emerge as a source of uncertainty that needs to be addressed \cite{Nobre:21, Perez:26}. This work demonstrates the ability of this framework to describe chromium neutron scattering and potentially many of other crucial reactions in this mass region \cite{Dupont:20}. 

More in general, establishing a microscopic description with an adequate description of neutron scattering in a relevant energy range, without it being tuned to any experimental data of the isotopes it investigates, provides a new tool to tackle fast neutron reaction problems. At the same time, this result demonstrates a pathway to extend predictive capability of microscopic reaction models to nuclei heavier than what previously considered feasible. Deriving the optical potential from a Hamiltonian, thereby directly linking reaction observables to underlying nuclear--structure constraints and properties, provides a new way to understand the nuclear system \cite{Idini:23proc}.

We have previously applied the method to calculate the spectra of $\isotope{48,49,50}{Cr}$ \cite{Ljungberg:22} showing good agreement with experiment for both energies and electromagnetic transitions (cf. Figure 4--10 of \cite{Ljungberg:22}).
In this article the framework is extended to reactions and it is applied to the scattering of neutrons on the nuclei $\isotope{48}{Cr}$ and $\isotope{50}{Cr}$ using the same SLy4-H Hamiltonian.
The calculation in this article was performed with $\Ncut=8$, $\qnum=10$, $\temp=0.45$, and the generator coordinates were sampled with $275$ points, giving $\ngrid=135$ and 152 GCM basis states below the cutoff $\MeV{25}$ for $\isotope{48}{Cr}$ and $\isotope{50}{Cr}$, respectively.
The resulting total cross sections are shown in \autoref{fig:tot_CS_Cr48_sig_a} and \autoref{fig:tot_CS_Cr50_sig_a} and compared to ENDF and JENDL nuclear data evaluations.

The cross section calculated using the GCM shows a clear improvement compared to the one given by the many--body mean--field $H_A$ only and it becomes comparable to the evaluated cross sections for energies between 2 and 10 MeV, where the present method is most applicable. The peaks at 8 and 11 MeV in the mean field cross sections are given by the $g_{7/2}$ and $h_{11/2}$ shells respectively.
It is notable that the lower energy peak is smeared out by the many--body correlations introduced in the GCM solution. This does not happen for the highest energy peak, where the GCM finds fewer states. 
This indicates the current high energy reach of our approach. 
In principle, this reach can be extended in a controlled manner by including single particle excitations in a more targeted way, satisfying more of Eq. \eqref{eq:almost_completeness}, as suggested in Appendix \ref{sec:quasiparticle}.
Furthermore, admitting more correlations and mixing states at high energy, e.g. by increasing the cutoff $E_\mathrm{cut}$ and the grid size $\ngrid$. We are implementing optimizations that are expected to yield substantial speed-ups, which will permit substantially larger $E_\mathrm{cut}$ and $\ngrid$ without prohibitive runtimes.

At energies below 2~MeV one can notice in \autoref{fig:tot_CS_Cr50_sig_a} the spread of the evaluated data points in the ENDF/B-VIII.0 dataset. 
The resolved resonant region, going from few keV to about 1 MeV in the case of $\isotope{50}{Cr}$, is characterized by a series of neutron $s$--wave neutron capture resonances that can be individually fitted by R--matrix. 
The subsequent energy range up to about 2 MeV is also dominated by the direct--semidirect neutron capture. Our model is presently not ideally suited for this region that is dominated by individual resonances. This is evidenced by the difficulty to converge at low energy, indicated by the difference between signature solutions. 
In Appendix \ref{sec:convergence} this problem is discussed in more detail together with possible improvements for this low--energy region.

The energy reach at both low and high energies can be extended by employing a Fock subspace tailored to the reaction observables of interest. The present calculation uses a Fock space designed to describe structure over a broad range of angular momenta. However, at low incident energies the neutron carries little orbital angular momentum and the cross section is dominated by $s$- and $p$-wave contributions. A basis optimized for these low-$\ell$ partial waves would admit more states and correlations for these partial waves, achieving improved accuracy at comparable computational cost.

\mkplot[0.48\textwidth]{plots}{tot_CS_Cr48_sig_a}{%
    Total neutron scattering cross section calculated for $\isotope{48}{Cr}$
        for $\sig=+i$, $a=\MeV{12}$ in solid orange, $\sig=-i$, $a=\MeV{12}$ in solid pink, and $\sig=+i$, $a=\MeV{24}$ in dashed purple,
    compared to JENDL-5 \cite{Iwamoto:23} in solid black and
    cross section calculated from the many--body mean--field potential $H_A$ only in dotted red.
}
\mkplot[0.48\textwidth]{plots}{tot_CS_Cr50_sig_a}{%
    Total neutron scattering cross section calculated for $\isotope{50}{Cr}$
    for $\sig=+i$, $a=\MeV{12}$ in solid yellow, $\sig=-i$, $a=\MeV{12}$ in solid pink, and $\sig=+i$, $a=\MeV{24}$ in dashed purple,
    compared to JENDL-5 \cite{Iwamoto:23} in solid black line,
    ENDF/B-VIII.0 as black points \cite{Brown:18}, and
    cross section calculated from the many--body mean--field potential only in dotted red.
}

Additionally, the differential elastic scattering cross section of $\isotope{50}{Cr}$ is compared to experimental and evaluated cross sections at 5 energies, shown in \autoref{fig:diff_CS_Cr50_sig_a}.
A compound nucleus contribution calculated using the semi-microscopic optical potention called JLM \cite{Bauge:01} was also added to the calculated differential cross sections in \autoref{fig:diff_CS_Cr50_sig_a}. 
In this energy range, mean field calculations tend to lack sufficient absorption and therefore overestimate the elastic channel, even though the positions of the diffraction minima are reasonably reproduced indicating corresponding reproduction of nuclear radii. By contrast, absolute differential cross sections computed within the GCM framework directly show excellent agreement with the evaluated data. Notably, doubling the strength of the imaginary term has only a minor impact on the elastic angular distribution at this energy. Furthermore, it is observable a discrepancy between experimental datasets: the angular distributions reported in~\cite{Korzh:75} at 2.5 and 3.0~MeV differ from those of~\cite{Fedorov:75} at 2.9~MeV. Only the latter are consistent with both our calculations and the ENDF/B‑VIII.0 evaluation.

\mkplot[0.48\textwidth]{plots}{diff_CS_Cr50_sig_a}{%
    Differential neutron scattering cross section calculated for $\isotope{50}{Cr}$
    for $\sig=+i$, $a=\MeV{12}$ in solid yellow, $\sig=-i$, $a=\MeV{12}$ in solid pink, and $\sig=+i$, $a=\MeV{24}$ in dashed purple,
    compared to experimental results as circles \cite{Korzh:75} and squares \cite{Fedorov:75},    ENDF/B-VIII.0 \cite{Brown:18} in dashdotted blue,
    and cross section calculated from the many--body mean--field potential only in dotted red.
    Laboratory energies of the neutron is shown,
    and angles shown are in laboratory system.
}

\section{Properties of the Optical Potential}  \label{sec:pot_props}

In what follows, we analyze the properties of our optical potential and benchmark them against other microscopic nonlocal optical potentials and the Koning--Delaroche global optical potential \cite{Koning:03}. This comparison illustrates the structure of the nonlocality and informs the design of next--generation phenomenological optical potentials~\cite{Hebborn:23}.

In general, the optical model potential is spherically symmetric, reflecting the underlying symmetries of the nuclear Hamiltonian, hence it can be defined for each partial wave by the decomposition
\begin{align*}
    V\of{\vec{r},\vec{r}'} & = \sum_{\ell m} V_\ell\of{r,r'} Y^\ell_m\of{\hat{r}} \conj{Y^\ell_m\of{\hat{r}'}} = \notag \\
    & = \sum_\ell \frac{2l+1}{4\pi} V_\ell\of{r,r'} P_\ell\of{\hat{r}\cdot\hat{r}'},
\end{align*}
where $Y^\ell_m$ are the spherical harmonics, $P_\ell$ are the Legendre polynomials, and $\hat{r} = \vec{r}/r$. In this work, the optical potential is constructed from wavefunctions, hence already separated in partial waves components  $V^{J\pi}(r,r',E)$, or equivalently $V^{\ell,J}(r,r',E)$, cf. Eq. \eqref{eq:finalgreens} and (\ref{eq:lippmann_schwinger}, \ref{eq:lippmann_schwinger_inv}). The partial wave components are often decomposed into a central part $V^\ell_c$, and a spin--orbit part $V^\ell_{ls}$ of the optical potential, according to
\begin{equation}
    V^\ell\of{r,r',E} = V^\ell_c\of{r,r',E} + \mathbf{L}\cdot \mathbf{S}~V^\ell_{ls}\of{r,r',E},
\end{equation}
where $\mathbf{L}$ and $\mathbf{S}$ are the angular momentum and spin operators of the incoming neutron. 
Therefore, the central part can then be easily calculated from the spin coupled potentials $V^{\ell,J}$ as
\begin{align}
    V^\ell_c\of{r,r',E} = & \frac{\ell+1}{2\ell+1} V^{\ell,\ell+\frac12}\of{r,r',E} + \notag \\
    & \frac{\ell}{2\ell+1} V^{\ell,\ell-\frac12}\of{r,r',E}.
\end{align}

Due to the energy--dependence of the optical potentials, the volume integrals per nucleon are often used to compare them. The volume integral is defined as
\begin{equation}
    J^\ell\of{E} = 4\pi \int \wrt{r} r^2 \int \wrt{r'} r'^2 V^\ell_c\of{r,r',E},
\end{equation}
and it has historically been used to fit and analyse numerous optical potentials,
such as \cite{Koning:03,Jeukenne:77, Idini:18} and refs. therein.

The real and imaginary parts of the volume integral are usually denoted $J^\ell_V\of{E} = \Re J^\ell\of{E}$ and $J^\ell_W\of{E} = \Im J^\ell\of{E}$, respectively.
In \autoref{fig:volint_real_cr50}, \ref{fig:volint_imag_cr50}, \ref{fig:volint_real_ncut10} and \ref{fig:volint_imag_ncut10}, the volume integrals per nucleon, $J^\ell_V\of{E}/A$ and $J^\ell_W\of{E}/A$, are shown for $\ell \le 3$ and for the nuclei $\isotope{50}{Cr}$ and $\mg{24}$.
$\mg{24}$ was the subject of \cite{Bostrom:25} and can be more easily compared to other studies in the literature \cite{Sargsyan:25,Rotureau:17}.
The $\mg{24}$ was calculated using $\Ncut=10$, $\temp=0.45$, $\sig=+i$ and had $\ngrid=179$ GCM basis states below the cutoff $\MeV{25}$.
The real part of the volume integral
varies between $-300\unit{MeV fm^3}$ and $-600\unit{MeV fm^3}$,
which is around the Koning Delaroche volume integrals at about $-450\unit{MeV fm^3}$,
though the variation is larger than
typically observed in phenomenological models \cite{Hodgson:84,Varner:91},
as well as in \cite{Sargsyan:25}
where a difference of only $5~\%$ is observed
between the different angular momenta $\ell$ for $\MeV{7.6}$.
The variations are larger for $\mg{24}$ and the $l=0$ partial wave of $\isotope{50}{Cr}$, while the other partial waves of $\isotope{50}{Cr}$ is quite close to the Koning Delaroche value.
Large variations indicate the presence of narrow resonances in the region involved,
and can be seen as well defined peaks in the imaginary part of the volume integral.
In the local optical model potential it is understood that the real part does not vary much
over a large range of different nuclei and projectile energies.
This is easily understood as the potential depth is expected to remain constant,
while the volume increases linearly with the number of nucleons $A$,
leaving the volume integral of the potential per nucleon roughly constant \cite{Hodgson:84,Varner:91}. However, in the case of a non-local model, the integration over the whole space could be influenced by the values of the potential for large non-locality $r-r'$ that will not influence observables as much as the part closer to the diagonal $r \approx r'$. 


\mkplot[0.48\textwidth]{plots}{volint_real_cr50}{%
    Real part of the volume integral of the central part of the total optical potential per nucleon for $\isotope{50}{Cr}$ fo selected $\ell$,
    compared to the real part of the volume integral of the global Koning Delaroche global optical potential \cite{Koning:03}.
}
\mkplot[0.48\textwidth]{plots}{volint_imag_cr50}{%
    Imaginary part of the volume integral of the central part of the total optical potential per nucleon for $\isotope{50}{Cr}$ fo selected $\ell$,
    compared to the imaginary part of the volume integral of the global Koning Delaroche global optical potential \cite{Koning:03}.
}
\mkplot[0.48\textwidth]{plots}{volint_real_ncut10}{%
    Real part of the volume integral of the central part of the total optical potential per nucleon for $\mg{24}$ fo selected $\ell$,
    compared to the real part of the volume integral of the global Koning Delaroche global optical potential \cite{Koning:03}.
}
\mkplot[0.48\textwidth]{plots}{volint_imag_ncut10}{%
    Imaginary part of the volume integral of the central part of the total optical potential per nucleon for $\mg{24}$ fo selected $\ell$,
    compared to the imaginary part of the volume integral of the global Koning Delaroche global optical potential \cite{Koning:03}.
}

The radial matrix elements for the real parts of
the calculated optical potentials are shown for $\mg{24}$ in
\autoref{fig:volint_real_radial_ncut10_0} for $l=0$ and
\autoref{fig:volint_real_radial_ncut10_2} for $l=2$,
and for $\isotope{50}{Cr}$ in \autoref{fig:volint_real_cr50} for $l=2$,
and were calculated using the same parameters as before
with an incoming neutron energy of $\MeV{3.25}$.
All exhibit similar features to the optical potentials
for the $s$- and $d_{3/2}$-waves of
$\isotope{16}{O}$ calculated in \cite{Rotureau:17},
using coupled cluster singles and doubles and $\text{NNLO}_\text{opt}$
nucleon-nucleon interaction \cite{Ekstrom:13}.
The similarities of the nonlocality despite differences
in both method, interaction, and nuclei
point towards this being a general feature,
at least for nuclei in this mass region.
The optical potential of the $d$-wave of $\isotope{50}{Cr}$
shows a similar nonlocality,
but lacks the repulsive part at low radii
seen in both \autoref{fig:volint_real_radial_ncut10_2}
and \cite{Rotureau:17}.
However, for low radii the scattering wavefunctions are
small, and so the behavior at low $r$ does not notably
affect the scattering observables,
as already discussed in \cite{Rotureau:17}.
It is important to note that a different basis size $\Nmax$
can impact the shape of the nonlocal potential.
For example, an eventual local component of the potential will slowly converge to a Dirac delta $\deltaf{\vec{r}-\vec{r}'}$ with increasing $\Nmax$.
$\Nmax=14$ was used in \cite{Rotureau:17}, while $\Nmax=16$ is used in this work, which allows for comparison of the potential shape.

The imaginary parts of the same partial waves are shown in \autoref{fig:volint_imag_radial_ncut10_0}, \ref{fig:volint_imag_radial_ncut10_2}, and \ref{fig:volint_imag_radial_cr50_2}.
The imaginary part extends to larger radii than the real part,
indicating that absorption is more important at the surface of the nucleus.

\mkplot[0.48\textwidth]{plots}{volint_real_radial_ncut10_0}{%
    Real part of the central $s$-wave optical potential of $\mg{24}$
    for an incoming neutron energy of $\MeV{3.25}$.
}
\mkplot[0.48\textwidth]{plots}{volint_real_radial_ncut10_2}{%
    Real part of the central $d$-wave optical potential of $\mg{24}$
    for an incoming neutron energy of $\MeV{3.25}$.
}
\mkplot[0.48\textwidth]{plots}{volint_real_radial_cr50_2}{%
    Real part of the central $d$-wave optical potential of $\isotope{50}{Cr}$
    for an incoming neutron energy of $\MeV{3.25}$.
}
\mkplot[0.48\textwidth]{plots}{volint_imag_radial_ncut10_0}{%
    Imaginary part of the central $s$-wave optical potential of $\mg{24}$
    for an incoming neutron energy of $\MeV{3.25}$.
}
\mkplot[0.48\textwidth]{plots}{volint_imag_radial_ncut10_2}{%
    Imaginary part of the central $d$-wave optical potential of $\mg{24}$
    for an incoming neutron energy of $\MeV{3.25}$.
}
\mkplot[0.48\textwidth]{plots}{volint_imag_radial_cr50_2}{%
    Imaginary part of the central $d$-wave optical potential of $\isotope{50}{Cr}$
    for an incoming neutron energy of $\MeV{3.25}$.
}

\section{Conclusion and Outlook} \label{sec:conclusion}

In conclusion, a general framework has been presented that computes neutron scattering cross sections directly from the spectra of a Hamiltonian. The approach remains applicable even when the calculated states do not form a complete basis, where the missing strength is estimated through sum rules introduced in~\cite{Bostrom:25} and further elaborated here, together with the interpretation of constructing an average generalized mean field.

In this work, the GCM with an effective Hamiltonian has been used to give a many-body solution. This was then applied to yield elastic scattering observables that are comparable to evaluated data. Neutron elastic scattering on ${}^{24}\mathrm{Mg}$, ${}^{48}\mathrm{Cr}$, and ${}^{50}\mathrm{Cr}$ has been calculated for both differential and total reaction cross section, obtaining good reproduction in the 2 to 10 MeV range without phenomenological adjustment. 
At these energies the present model provides a significant improvement over a spherical mean–field potential constructed from the same interaction. In particular ${}^{50}\mathrm{Cr}$, a key isotope for applications has been investigated.

The calculation provides additional insight into properties of the optical potential, including its nonlocal character, and suggests properties useful for the development of future phenomenological potentials. In this direction, the potential and its functional form can be parametrized through an orthogonal basis expansion~\cite{Nord:thesis}, enabling systematic surveys along isotopic chains and across additional mass regions.

The present implementation also identifies the degrees of freedom that should be prioritized when tailoring a Fock subspace for reaction observables. At low incident energies, where $s$- and $p$-wave contributions dominate, a generator coordinate basis optimized for low-$\ell$ can admit more correlations at comparable computational cost. More broadly, the results indicate that cross sections are more sensitive to correlations and harder to converge than Yrast spectra, frequently requiring a denser set of basis states optimized for the physically relevant partial waves.

Several extensions are underway and have been identified in this manuscript. In particular, spectroscopic amplitudes and sum rules are being developed to compute excited target states, which will enable an explicit treatment of multichannel inelastic scattering within a $J$–matrix formalism~\cite{Bostrom:26b}.
These upgrades are expected to further strengthen the link between structure and reaction observables and to improve the reliability of the cross sections approaching the resonant region below $\sim 2$~MeV.

Because the effective Hamiltonian is rooted in DFT and is not additionally adjusted with isotope–specific parameter tuning, the method is applicable to heavier nuclei and to nuclei far from stability. In this context, the framework provides a practical bridge between nuclear structure and nuclear reactions. Within a model that reproduces elastic scattering data, it becomes possible to identify which components of the Hamiltonian and which structure observables correlate most strongly with specific reaction observables potentially further constraining nuclear theory and enhancing the ability to generate microscopic data needed for nuclear technology and astrophysical applications.

\section*{Acknowledgements}
This work has been supported by the Swedish Research Council (Vetenskapsr{\aa}det) VR~2020-03721, Knut and Alice Wallenberg foundation (KAW~2015.0021), Crafoord foundation, and Krapperup foundation. Computing was enabled by resources provided by the National Academic Infrastructure for Supercomputing in Sweden (NAISS), partially funded by the Swedish Research Council through grant agreement no. 2022-06725, and The Centre for Scientific and Technical Computing at Lund University (LUNARC).


\bibliographystyle{apsrev4-2}
\bibliography{sources.bib,nuclear.bib}

\appendix

\section{Convergence}\label{sec:convergence}

We have recently applied the method to describe neutron scattering on $\mg{24}$ \cite{Bostrom:25}. This nucleus is here used as a test case to study the convergence characteristics of the method.
$\mg{24}$ was chosen since it is light but deformed,
as well as being well studied in previous literature,
and it's low energy spectra was previously
calculated using the GCM and compared well to experiment \cite{Ljungberg:22}.
The convergence properties of the method were studied with $\Nmax=16$ and $\Ncut=6$,
varying $\ngrid$, $\temp$, $\qnum$, and the chosen signature $\sig$.
While $\Ncut=6$ is not large enough to achieve convergence,
it still gives a representative image of
how each convergence parameter affects the result.
The parameters of the Hamiltonian was held constant
for all calculations on Mg, with the value of
$\chi = 5.817\times 10^{-3}\,\unit{MeV}^{-1}$
obtained from a $\Ncut=10$ fit.

The HO basis size $\Nmax$ was first investigated,
and $\Nmax=16$ was chosen for the subsequent calculations,
then the size of the truncated single particle basis $\ncut$ was investigated.
Then, the number of HFB vacua $\ngrid$, the temperature $\temp$, the number of quasiparticle excitations to choose from $\qnum$, and the signature $\sig$, were varied independently for $\Ncut=6$.
The energy cutoff $E_\mathrm{cut}$ was chosen as $\MeV{25}$.

\subsection{N{max} and Regularization}

In \autoref{fig:tot_CS_static}, a comparison between
different basis sizes \Nmax{} is shown, and the effect of regularization,
using the cross section calculated using only the spherical reference potential $\op{H}_0$ given by HOSPHE for different \Nmax{}.
The oscillations caused by the Gibbs phenomenon create additional unphysical scattering,
leading to a higher total cross section.
As seen in \autoref{fig:tot_CS_static},
this effect is significant even at $\Nmax=32$.
However, when using the regularization, the Gibbs phenomenon is reduced,
and the cross section does not change significantly when changing \Nmax{}.
This shows that the regularized $\Nmax=16$ is
very close to achieving convergence.
The large cross section with $\Nmax=32$
without regularization is purely an artifact of the finite basis.
In addition, the fact that the smoothing factors
\eqref{eq:smooth_sigma} converge to 1 as \Nmax{} increases
ensures that the cross section converges
to the exact cross section in that limit.
This can be understood physically,
considering that the potential matrix elements are significantly
altered by the regularization only for high harmonic oscillator quanta.
These matrix elements should not contribute significantly to the scattering of low energy particles.
For the $\Nmax=32$ calculation, the parameter $\alpha$ was varied between
$\alpha=3$ to $\alpha=10$ without visible differences
in the resulting cross sections,
showing the calculation is insensitive to the exact value of the smoothing parameter $\alpha$.

There is a notable peak in the cross section around $\MeV{1}$,
which was identified as originating from the $f_{7/2}$ partial wave.
This indicates the existence of a long lived resonance state
in the spherical reference potential with an energy near $\MeV{1}$.
However, resonances appearing in the spherical reference potential
are in general not preserved when considering correlation and calculating the full many--body optical potential.

\mkplot[0.48\textwidth]{plots}{tot_CS_static}{%
    Total neutron scattering cross section of the spherical reference potential of $\mg{24}$
    calculated for different values of \Nmax{},
    compared to a cross section calculated without regularization.
    The solid red line corresponds to $\Nmax=16$
    and the dotted blue line corresponds to $\Nmax=32$,
    both calculated using regularization,
    and the dashed yellow line corresponds to $\Nmax=32$
    but without using regularization.
}

\subsection{\Ncut}

In \autoref{fig:tot_CS_Ncut}, the calculated total cross sections
for different values of \Ncut{} are shown.
All calculations used $\qnum=10$, $\temp=0.45$, $\sig=+i$,
and sampled the generator coordinates with 275 points,
giving $\ngrid=207$, 186, and 179 GCM basis states below the cutoff $\MeV{25}$
for $\Ncut=6$, $\Ncut=8$, and $\Ncut=10$ respectively.
Our current implementation was used to calculate up to $\Ncut=10$ for $\isotope{24}{Mg}$
within reasonable time, and this will be improved in future versions.
There is a noticeable difference between $\Ncut=8$ and $\Ncut=10$ especially
at lower energies, showing a need for larger $\Ncut$ to achieve full convergence in this case.
Though, $\Ncut=10$ gives a cross section that is only on average about $2.1~\%$ smaller than the cross section calculated for $\Ncut=8$, a difference not significant to compare to experiment and within the error bounds provided (cf. fig. \ref{fig:tot_CS_Ncut}).

Despite the $\Ncut=6$ not being converged with respect to \Ncut{},
it was still used to perform tests of other convergence parameters,
in order to test a wide variety of parameters.
Since convergence can still be achieved within the limited
space described by the smaller single particle basis,
it still showcases the properties of the method,
and gives guidance on how to choose parameters for
calculations using larger values of $\Ncut$, for which a full variation
of parameters is not feasible.

\mkplot[0.48\textwidth]{plots}{tot_CS_Ncut}{%
    Total neutron scattering cross section for $\mg{24}$ calculated for different values of \Ncut{}.
    The dashed pink line corresponds to $\Ncut=6$,
    the dotted blue line corresponds to $\Ncut=8$,
    and the solid green line corresponds to $\Ncut=10$.
}

\subsection{Number of vacua}

In \autoref{fig:tot_CS_grid_edited}, the calculated total cross sections
for different values of $\ngrid$ are shown.
All calculations used $\Ncut=6$, $\qnum=10$, $\temp=0.45$, $\sig=+i$,
and they sampled the generator coordinates with
70, 150, 200, 275, and 700 points,
giving $\ngrid=54$, 112, 145, 207, and 507 GCM basis states below the cutoff $\MeV{25}$ respectively.
As the number of vacua increases, a larger energy region can be seen to converge, with $\ngrid\geq112$ looking converged above $\MeV{8}$,
while $\ngrid\geq145$ gives convergence above around $\MeV{5}$.
Finally, with $\ngrid=207$ the energy range above $\MeV{3}$ looks converged to the cross sections for $\ngrid=507$.
$\ngrid=507$ only gives a cross section on average $6.0\unit{\%}$ larger than $\ngrid=207$, and $2.7\unit{\%}$ larger above $\MeV{3}$.

At low energies, the density of states is lower,
and so fewer states are mixed to form narrow resonances, establishing the resolved resonance and the direct--semidirect neutron capture regions. In this resonant energy region, the cross sections will be more sensitive to the precise
distribution of energy and spectroscopic strength between the states and partial waves. This makes the calculation difficult to converge, as small differences in the calculation will reflect strongly for the cross section, without explicitly accounting for the many-body coupling to the continuum. 
For higher energies, these details are less important, and more states over a wider energy range contribute to the resonances,
making the cross section more robust to changes to the individual states that make up the resonance.
In addition, the completion ensures the resonances are given the correct energy centroids and total strength,
which also decrease the relative importance of any single state.
Because of this, convergence is more difficult for low projectile energies, and increasing the basis progressively improves convergence in that region.

Specifically, the large difference at low energy is mainly due to the $p_{1/2}$ partial wave, where a low-energy resonance makes it difficult to reach convergence without explicitly coupling the many-body $1/2^-$ state to the continuum.


\mkplot[0.48\textwidth]{plots}{tot_CS_grid_edited}{%
    Total neutron scattering cross section for $\mg{24}$ calculated for different values of \ngrid{}.
    The red dash--dot--dotted line corresponds to $\ngrid=54$,
    the purple dash--dotted line corresponds to $\ngrid=112$,
    the green dotted line corresponds to $\ngrid=145$,
    the pink dashed line corresponds to $\ngrid=207$,
    and the solid blue line corresponds to $\ngrid=507$.
}

\subsection{Temperature}

In \cite{Ljungberg:22}, the optimal range of the \temp{} parameter
was previously found to be $\temp=[0.3,0.6]$,
so to investigate how \temp{} affect the cross section,
calculations were performed for $\temp=0.3$, $\temp=0.45$, and $\temp=0.6$.
In \autoref{fig:tot_CS_T}, the calculated total cross sections
for these different values of $\temp$ are shown.
All calculations used $\Ncut=6$, $\qnum=10$, $\sig=+i$,
and sampled the generator coordinates with 275 points,
giving $\ngrid=207$ GCM basis states below the cutoff $\MeV{25}$
for $\temp=0.45$ and $\temp=0.6$, and $\ngrid=206$ for $\temp=0.3$.
These calculations are compared to the result obtained with a large number of vacua $\ngrid = 507$, which is the most well converged calculation in the model space.
At higher energies, the four results are quite similar,
though the cross section for $\temp=0.6$ is slightly lower.
It is only for energies in the resonant region that significant differences appear. As explained in the previous section, it is expected to have difficulty converging within this energy range.

A larger $\temp$ introduces more particle--hole excitations in the HFB basis states, which should improve the convergence of states with a large single--particle characteristic, which are the ones with large spectroscopic factors and therefore contribute the most to scattering.
However, the temperature procedure also introduces multi particle--multi hole excitations,
due to the exponential nature of the operator,
which has the opposite effect on the convergence of states with large spectroscopic factor. 

Since $\temp=0.45$ was found to give best convergence of energy levels in \cite{Ljungberg:22}, and to reproduce the $\ngrid =507$ results with slightly better agreement at low energy than $\temp=0.3$, this value was also used for the calculations in this article. In \autoref{fig:tot_CS_T}, the low energy cross section appear more sensitive to the temperature, and a larger number of GCM basis states would be necessary for the results to become independent of $\temp$.

\mkplot[0.48\textwidth]{plots}{tot_CS_T}{%
    Total neutron scattering cross section for $\mg{24}$ calculated for different values of \temp{}.
    The dotted yellow line corresponds to $\temp=0.6$,
    the solid pink line corresponds to $\temp=0.45$,
    and the dashed red line corresponds to $\temp=0.3$. These are compared to the calculation with the largest number of vacua $\ngrid = 507$ in dashed dotted blue line.
}

\subsection{Signature}

When choosing the quasiparticle to excite
in order to construct the odd particle number state,
one has a choice of exciting a quasiparticle of signature
$\sig=+i$ or $\sig=-i$.
Since the Hamiltonian is rotationally invariant,
both signatures should converge to the same solution.
This fact makes it possible to use the solutions
for the two different signatures to estimate a degree of convergence.

To compare the two signature choices,
calculations were performed using $\Ncut=10$, $\qnum=10$, $\temp=0.45$,
and two different $\ngrid$.
In \autoref{fig:spectra_sig_Ncut6_minigrid_25+},
the spectra calculated using different signature quasiparticles
are compared for a calculation using only $\ngrid=54$ GCM basis states,
showing large differences between the two signature choices.
Using instead $\ngrid=179$ GCM basis states below the cutoff $\MeV{25}$,
both signature choices converges towards the same spectra,
as can be seen in \autoref{fig:spectra_sig_25+},
which shows a good convergence of the yrast band.
There are larger differences in the excited states,
though it is important to remember that most of these
have small spectroscopic factors and so individually contribute
very little to the overall cross section.

\mkplot[0.48\textwidth]{plots}{spectra_sig_Ncut6_minigrid_25+}{%
    Spectrum of positive parity states of $\mg{25}$ calculated for different signatures \sig{} for $\Ncut=6$ and $\ngrid=54$,
    showing the result of an unconverged calculation.
    The blue circles corresponds to $\sig=+i$ and
    the red stars corresponds to $\sig=-i$.
    All energies are relative to the lowest ground state energy.
    The calculated threshold energy is shown as a red dotted line.
}
\mkplot[0.48\textwidth]{plots}{spectra_sig_25+}{%
    Spectrum of positive parity states of $\mg{25}$ calculated for different signatures \sig{} with $\Ncut=10$ and $\ngrid=179$.
    The green circles corresponds to $\sig=+i$ and
    the blue stars corresponds to $\sig=-i$.
    All energies are relative to the lowest ground state energy.
    The calculated threshold energy is shown as a red dotted line.
}

The cross sections of the two
signature choices are compared in \autoref{fig:tot_CS_bigsig}
for the $\ngrid=179$ case.
For intermediate energies, they correspond quite well,
while the difference between them increases for low energies,
convergence is more difficult.

\mkplot[0.48\textwidth]{plots}{tot_CS_bigsig}{%
    Total neutron scattering cross section for $\mg{24}$ calculated for different signatures \sig{}.
    The solid green line corresponds to $\sig=+i$
    and the dotted blue line corresponds to $\sig=-i$.
    Cross sections calculated using the Koning--Delaroche optical potential \cite{Koning:03} is included for comparison as a gray double--dotted line.
}

It is to be noted that the spectra calculated with
this particular Hamiltonian reproduces
the low energy bands of $\mg{25}$ but predicts the ground state
to have spin $1/2$, when experimental data shows
the ground state of $\mg{25}$ to have spin $5/2$.
This discrepancy can be traced to the deformation dependence of
the energies of the valence shell levels $1s_{1/2}$ and $0d_{5/2}$.
As shown in \autoref{fig:nilsson_n_edited2}, the
$d_{5/2}$ $K=5/2$ orbital increases in energy when the deformation increases,
crossing the $s_{1/2}$ $K=1/2$ orbital close to the ground state deformations
of $\mg{24}$ and $\mg{25}$, and so the relative energies of the
bands are very sensitive to the exact deformation of the nucleus.

\mkplot[0.48\textwidth]{plots}{nilsson_n_edited2}{%
    Single-particle neutron energies calculated for $\mg{24}$ as a function of the deformation $\beta = \sqrt{\beta_x^2 + \beta_y^2}$, showing the valence $s$-$d$-shell,
    calculated using $\Nmax=16$ and $\Ncut=10$.
    The $d$-shells are shown with solid lines, the $s$-shell is shown with a dotted line, and the $f$-shell is shown with dashed lines.
    The magic numbers 8 and 20 are shown at their respective shell closures. 
}

The fact that $I=1/2$ state achieves a lower energy than the $I=5/2$ state indicates that the deformation stiffness is slightly too low.
Since a higher deformation effectively means the nucleus will have a slightly larger spatial extent, this could in turn result in a larger cross section, which could explain the systematically too large cross sections compared to experiment shown in \autoref{fig:tot_CS_Ncut}.
This shortcoming does not invalidate the conclusions derived from these results, as the convergence characteristics will still be representative of the method,
though the Hamiltonian could be tuned to reproduce the experimental yrast band or other observables instead of being exclusively generated from DFT.

\subsection{Quasiparticle Excitation}

The even--odd GCM basis is constructed by randomly
choosing and excitating a quasiparticle
among the \qnum{} lowest energy quasiparticle states.
Exciting only the lowest energy quasiparticle for each vacuum ($\qnum = 1$) would predominantly select single--particle excitations in the valence shell, allowing the selection of different single--particles only through the deformation and temperature operator. On the other hand, a large \qnum{} will allow the GCM procedure to directly
sample a larger space of single--particle excitations.
However, exciting high energy quasiparticles
is detrimental to the description of low energy phenomena,
so a too large value of \qnum{} with respect to \ngrid{} will result in too few
basis states that can adequately describe the low energy
structure and scattering.
Due to the variational principle, the true yrast band should have
lower energy than the calculated states,
and so a lower energy yrast state indicates a more converged state. 

In \autoref{fig:tot_CS_qnum},
the calculated total cross sections
for different values of \qnum{} are shown.
All calculations shown in Figures \ref{fig:tot_CS_qnum}--\ref{fig:qp_csum} used $\Ncut=6$, $\temp=0.45$, $\sig=+i$,
and they sampled the generator coordinates with 275 points,
giving $\ngrid=207$ GCM basis states below the cutoff $\MeV{25}$. These calculations are also compared to the result obtained with a large number of vacua $\ngrid = 507$. $\qnum = 10$ agrees best with the latter above $\MeV{3}$.

As seen in \autoref{fig:spectra_qp_25+}, $\qnum=10$ and $\qnum=5$
gives the lowest energies for the yrast band.
This is due to the fact that the yrast band is mostly a collective
band with the odd particle in the valence shell, $s_{1/2}$ or $d_{5/2}$.
Including higher energy excitations in the basis ($\qnum = 15$) leaves fewer basis states to describe the yrast band.

\mkplot[0.48\textwidth]{plots}{tot_CS_qnum}{%
    Total neutron scattering cross section for $\mg{24}$ calculated for different number of quasiparticle excitations \qnum{}.
    The dashed blue line corresponds to $\qnum=5$,
    the solid pink line corresponds to $\qnum=10$,
    and the dotted green line corresponds to $\qnum=15$. These are compared to the calculation with the largest number of vacua $\ngrid = 507$ in dashed dotted blue line.
}

\mkplot[0.48\textwidth]{plots}{spectra_qp_25+}{%
    Spectrum of positive parity states of $\mg{25}$ calculated for different number of quasiparticle excitations \qnum{}.
    The blue circles corresponds to $\qnum=5$,
    the light red 4--pointed stars corresponds to $\qnum=10$,
    and the green 5--pointed stars corresponds to $\qnum=15$.
    All energies are relative to the lowest ground state energy.
    The threshold energy is shown as a red dotted line.
}

At higher energies, it's harder to judge
how well the calculation performs from the spectra alone.
\autoref{fig:qp_csum} shows the cumulative sum of spectroscopic factors
up to a given energy, counted with degeneracy, for spherical $\op{H}_0$ compared to the GCM solution for different $\qnum$ without completion of the sum rule.
Since the terms in the Green's function expansion is proportional to
the spectroscopic factors, this also shows how the contributions to
the Green's function and, by extension, the optical potential,
is distributed in energy.
The spherical reference $\op{H}_0$ eigenstates show clear discontinuities at each single--particle energy level,
the size of each being the degeneracy of the spherical shell,
since the spectroscopic factors of the levels are all exactly one.
All three GCM calculations can be seen to agree quite well up to about $\MeV{5}$,
but above this energy $\qnum=5$ increases slowly,
indicating that few states contributing to
the optical potential are found above this energy.
In contrast, the calculations with $\qnum=10$ and 15 continue finding
states up to about $\MeV{15}$.
This shows that including more higher energy excitations is helpful for
finding the states contributing to the optical potential at higher energies.
In conclusion, $\qnum=10$ shows good convergence for both low and high
energy states, corresponds best to the large number of vacua calculations in the relevant region, and performs better overall than $\qnum=5$ or $\qnum=15$.

\mkplot[0.48\textwidth]{plots}{qp_csum}{%
    Sum of all spectroscopic factors below the energy $E$, counted with degeneracy, calculated for different number of quasiparticle excitations \qnum{}.
    The blue solid line corresponds to $\qnum=5$,
    the pink dotted line corresponds to $\qnum=10$,
    the green dashed line corresponds to $\qnum=15$,
    and the red double--dotted line corresponds to the spherical reference potential single--particle states.
}

\section{Effect of completion on cross sections}\label{sec:completion}



In \autoref{fig:biggrid_sf2_combined_edited_nog},
the spectroscopic functions with and without completion are shown
and compared to Lorentz-broadened single-particle levels
for a calculation with $\Ncut=6$, $\temp=0.45$, $\sig=+i$,
and 275 points that sample the generator coordinates,
giving $\ngrid=207$ GCM basis states below the cutoff $\MeV{25}$.
For energies around the Fermi energy $\MeV{-12.78}$,
the uncompleted and completed states give
the same value of the spectroscopic function,
indicating that the GCM without completion finds a majority of the states with significant contributions.
This shows that the method can be expected to give contributions
in the energy interval $E_F \pm E_\mathrm{cut}$.
Since the Fermi energy corresponds roughly to the
separation energy of the scattering particle,
the method is predicted to perform best for neutron rich nuclei,
as then the center of the energy range of validity
would be closer to positive energies, where the scattering occur.

In \autoref{fig:tot_CS_biggrid_24Mg_A+1} the effect of the many--body mean--field correlations on cross sections is investigated. Neutron scatttering is computed using the completion procedure in three ways: i) using the full many--body mean--field $H_A$ as Eq. \eqref{eq:energy_sumrule}; ii) using the spherical reference potential $H_0$ instead on the left--hand side of Eq. \eqref{eq:energy_sumrule}, neglecting the contribution of the two--body part of the Hamiltonian; iii) using $H_A$ but without using hole states from the GCM and relying only on completion to account for $A-1$ contribution. The three cross sections only differ significantly in the resonant region. While the exact expression is important for consistency, neglecting the two--body contribution to the mean--field is, in this case, a reasonable approximation for higher energies.

The hole states are far in energy from the projectile energies, as in \autoref{fig:biggrid_sf2_combined_edited_nog}. However, their presence is needed to guarantee the correct properties of the Green's function and self energy. While it is possible to directly include holes from a reference potential in a closed shell case, this is not straightforward in the open-shell case. The completion procedure guarantees the correct energy centroid and spectroscopic strength so that one can avoid the explicit many-body calculation of the $A-1$ system in this case, with satisfactory results above the resonant region.

\mkplot[0.48\textwidth]{plots}{tot_CS_biggrid_24Mg_A+1}{%
    Total neutron scattering cross section for $\mg{24}$ calculated when completing the many--body basis with different potentials.
    The solid pink line corresponds to using the full calculation using the exact expression for the many--body mean--field $H_A$ and both $A+1$ and $A-1$ states for Eq. \eqref{eq:energy_sumrule},
    the dashed blue line corresponds to using only $A+1$ states for Eq. \eqref{eq:energy_sumrule},
    the dotted line corresponds to using the spherical reference potential $H_0$ instead.
}

\section{Imaginary part}\label{sec:imag_part}

The imaginary parameter $i\eta$ in the K\"all\'en--Lehmann representation of the Green's function \eqref{eq:finalgreens} originates from the Fourier transform of the step function indicating the propagation of a particle backward and forward in time. Formally, it should be taken to zero, to enforce causal boundary conditions in the particle propagation. In this framework, by constructing the self energy $\Sigma$ by inverting the Green's functions $G$ and $G^0$ \eqref{eq:sigma}, the imaginary part of the Green's function is directly related to the imaginary part of the optical potential $\Im \Sigma$, which in turn incorporates the dissipation of the elastic scattering cross section into other reaction channels, impacting both the elastic and the total scattering cross section. While the underlying dynamics is time--reversal invariant, the projected optical Hamiltonian, i.e. effective representation of the scattering process through a complex potential, is not necessarily symmetric with respect to time: an incoming neutron can be absorbed or scatter inelastically and its energy is then dissipated in other channels that are not elastic scattering. Therefore, the imaginary part of the Green's function denominator, hence of the microscopic optical potential, should account for this dissipation effect to channels outside the elastic subspace that the nuclear Hamiltonian is projected to.
Some channels can be accounted for in consistent ways by means of Berggren basis \cite{Berggren:68,Hebborn:23}, Gamow shell model \cite{Michel:21}, or Green's function consistency \cite{Sargsyan:25}, obtaining complex energies of specific resonant and quasibound states.
Electromagnetic transitions can be accounted for using time-dependent perturbation theory, giving an additional contribution to the imaginary part of the energy through Fermi's golden rule.

However, in the present implementation of this framework, the imaginary part is exclusively provided by $i\eta$ in the denominator. Consequently, strictly taking the limit $\eta \rightarrow 0^+$ will make $\Im \Sigma$ go to zero almost everywhere. Therefore, the dissipation effects will have to be effectively represented by a finite value of $\eta$. Despite the finite value of $i\eta$, the dispersive Kramers-K\"oning relation is satisfied by construction, together with causality.

Additionally, a many--body method will be hard pressed to reproduce the density of states through eigenstates of an Hamiltonian, and any computational method will fail at several MeV energy, even in relatively light nuclei close to shell closure such as the chromium and magnesium isotopes here analyzed. In principle, these eigenstates can be coupled to the continuum and will form a continuous strength distribution. However, the details of every single state and its specific coupling to the continuum are irrelevant, it is the total strength distribution for each partial waves that determines the phase shift and the reaction observables. An imaginary component spreads energy poles in an energy range following a Lorentzian distribution as in \autoref{fig:biggrid_sf2_combined_edited_nog}. Even if a many-body method is not explicitely coupled to the continuum, it can reliably reproduce the strength distribution, provided a nominal width $\eta$ is associated with each state. Therefore, finite value of $\eta$ has also the role to numerically stabilize the scattering solution by averaging out the spectrum distribution, avoiding unphysical peaks. If the spectrum obtained is dense enough, the obtained distribution will be largely independent of $\eta$ because its average will not be affected by the width of the Lorentzian folding.

In our present implementation, the dissipation component of $\Im \Sigma$ and the spectrum averaging are both parametrized by the factor $i \eta$ in \eqref{eq:finalgreens} in a phenomenological way by $\frac{a}{\pi}\frac{(E-E_\mathrm{F})^2}{(E-E_\mathrm{F})^2 + b_\eta^2}$. Here,
the dependence on the choice of $a$, i.e. the magnitude of $\eta$, is investigated. 
The total cross sections calculated for three different values of $a = 6, 12, 24$~MeV are shown in \autoref{fig:tot_CS_biga}, for neutron scattering on $\isotope{24}{Mg}$.
These calculations used $\Ncut=10$, $\qnum=10$, $\temp=0.45$, $\sig=+i$, 
and sampled the generator coordinates with 275 points,
giving $\ngrid=179$ GCM basis states below the cutoff $\MeV{25}$.
In \autoref{fig:tot_CS_biga}, one can see that using a larger imaginary part results in slightly lower cross sections.
Increasing $a$ by a factor of 4 from $\MeV{6}$ to $\MeV{24}$
gives an average decrease of the cross section by approximately $9~\%$,
much smaller than the variation in $a$. Similarly, the differential elastic scattering cross section in \autoref{fig:diff_CS_Cr50_sig_a} is marginally affected by the doubling of $a$.

We conclude that the total cross section
is quite insensitive to the magnitude of the imaginary part
over a range of values of $a$.
This also implies that the total cross section
would be on average similarly insensitive to
more consistent prescriptions
for including the effects of the continuum, especially at energies above the resonant region. However, in \autoref{fig:tot_CS_biga} it is clear that the resonant part of the spectrum is more affected by the choice of $a$ than the higher energy region, indicating that including explicit continuum coupling could improve the reliability and convergence properties of that lower energy region.

\mkplot[0.48\textwidth]{plots}{tot_CS_biga}{%
    Total neutron scattering cross section for $\mg{24}$ calculated for different values of $a$ in $\eta = \frac{a}{\pi}\frac{(E-E_\mathrm{F})^2}{(E-E_\mathrm{F})^2 + b_\eta^2}$.
    The green dotted line corresponds to $a=\MeV{6}$,
    the blue solid line corresponds to $a=\MeV{12}$,
    and the dashed purple line corresponds to $a=\MeV{24}$.
}

\section{Quasiparticle excitations}
\label{sec:quasiparticle}
For each even-even basis state, only one quasiparticle is excited and used to construct the odd-even basis.
This is due to the fact that the Hamiltonian is obtained
by calculating $\frac{\braopket{\phi_a}{H}{\phi_b}}{\braket{\phi_a}{\phi_b}}$
and $\braket{\phi_a}{\phi_b}$ separately, before combining them as
\begin{equation}
    \braopket{\phi_a}{H}{\phi_b} =
    \frac{\braopket{\phi_a}{H}{\phi_b}}{\braket{\phi_a}{\phi_b}} \braket{\phi_a}{\phi_b}.
\end{equation}
But since two different quasiparticle excitations on the same vacuum
are orthogonal, this decomposition breaks down since
\begin{equation}
    \frac{\braopket{\phi_a}{H}{\phi_b}}{\braket{\phi_a}{\phi_b}}
\end{equation}
diverges.
Some ways to avoid this problem exists \cite{Donau:98}.
A possibility within the present framework is choosing appropriate superpositions of excitations,
such as $\hconj{\beta}_1\ket{\phi}$ and
$\frac{1}{\sqrt{2}}\left(\hconj{\beta}_1 + \hconj{\beta}_2\right)\ket{\phi}$,
as these span the same space as $\hconj{\beta}_1\ket{\phi}$ and $\hconj{\beta}_2\ket{\phi}$,
but are not orthogonal.

\end{document}